\documentclass[prd, twocolumn, superscriptaddress,floatfix, nofootinbib]{revtex4}

\usepackage{url}
\usepackage{xspace}
\usepackage{dsfont}
\usepackage{amssymb}
\usepackage{amsmath}
\usepackage{graphicx}
\usepackage[small]{subfigure}
\usepackage[colorlinks=true,citecolor=blue]{hyperref}
\usepackage{multirow}
\usepackage{hhline}

\usepackage{color}
\definecolor{darkred}{rgb}{1.0,0.1,0.1}
\definecolor{darkgreen}{rgb}{0.1,0.7,0.1}
\definecolor{darkblue}{rgb}{0.1,0.1,1.0}

\newcommand{\BN}[1]{\textbf{\color{red}[#1 --BN]}}
\newcommand{\DS}[1]{\textbf{\color{orange}[#1 --DS]}}

\usepackage{amsmath}
\DeclareMathOperator*{\argmax}{argmax}
\DeclareMathOperator*{\argmin}{argmin}

\def\beq{\begin{equation}}
\def\eeq{\end{equation}}
\newcommand{\bea}{\begin{eqnarray}\begin{aligned}}
\newcommand{\eea}{\end{aligned}\end{eqnarray}}

\begin{document}

\title{DCTRGAN: Improving the Precision of Generative Models with Reweighting}

\author{Sascha Diefenbacher}
\email{sascha.daniel.diefenbacher@uni-hamburg.de}
\affiliation{Institute f\"{u}r Experimentalphysik, Universit\"{a}t Hamburg, Germany}

\author{Engin Eren}
\email{engin.eren@desy.de}
\affiliation{Deutsches Elektronen-Synchrotron, Germany}

\author{Gregor Kasieczka}
\email{gregor.kasieczka@uni-hamburg.de}
\affiliation{Institute f\"{u}r Experimentalphysik, Universit\"{a}t Hamburg, Germany}

\author{Anatolii Korol}
\email{anatolii.korol@desy.de}
\affiliation{Taras Shevchenko National University of Kyiv, Ukraine}

\author{Benjamin Nachman}
\email{bpnachman@lbl.gov}
\affiliation{Physics Division, Lawrence Berkeley National Laboratory, Berkeley, CA 94720, USA}

\author{David Shih}
\email{shih@physics.rutgers.edu}
\affiliation{NHETC, Department of Physics and Astronomy, Rutgers University, Piscataway, NJ 08854, USA}

\begin{abstract}
Significant advances in deep learning have led to more widely used and precise neural network-based generative models such as Generative Adversarial Networks (\textsc{Gans}).  We introduce a post-hoc correction to deep generative models to further improve their fidelity, based on the \textit{Deep neural networks using the Classification for Tuning and Reweighting} (\textsc{Dctr}) protocol. 
The correction takes the form of a reweighting function that can be applied to generated examples when making predictions from the simulation.
We illustrate this approach using \textsc{Gans} trained on 
standard multimodal probability densities as well as calorimeter simulations from high energy physics. We show that the weighted \textsc{Gan} examples significantly improve the accuracy of the generated samples without a large loss in statistical power.  This approach could be applied to any generative model and is a promising refinement method for high energy physics applications and beyond.   
\end{abstract}

\maketitle

\section{Introduction}

Generative models are a critical component of inference in many areas of science and industry.  As a result of recent advances in deep learning, neural network-based generative models are rapidly being deployed to augment slow simulations, act as simulator surrogates, or used for ab initio modeling of data densities directly for inference or for uncertainty quantification.  Well-studied methods include Generative Adversarial Networks (\textsc{Gan})~\cite{Goodfellow:2014:GAN:2969033.2969125,Creswell2018}, Variational Autoencoders~\cite{kingma2014autoencoding,Kingma2019}, and variations of Normalizing Flows~\cite{10.5555/3045118.3045281,Kobyzev2020}.

In many industrial applications of generative modeling, the aim is primarily to achieve ``realistic" images on a per-example basis. By contrast, in high energy physics applications, the main goal is usually to improve the quality of an {\it ensemble of examples}. In other words, it is of paramount importance that the generator accurately reproduce the underlying probability distribution of the training data. 

One challenge faced by current generative models is that even though they are able to qualitatively reproduce features of the data probability density, they are often unable to reproduce fine details.  While a variety of advanced methods are being proposed to enhance the precision of generative models, we make the observation that a relatively simple operation can be performed on the output of a generative model to improve its fidelity.  This operation uses a neural network classifier to learn a weighting function that is applied post-hoc to tweak the learned probability density.  The result is a set of weighted examples that can be combined to more accurately model the statistics of interest.  This procedure does not improve the fidelity of any particular example, but instead improves the modeling of the probability density.

Estimating probability density ratios with classification has a long history (see e.g., Ref.~\cite{hastie01statisticallearning,sugiyama_suzuki_kanamori_2012}) and has been used for a wide variety of applications in machine learning research related to generative models~\cite{10.5555/3157096.3157127,mohamed2016learning,grover2017boosted,NIPS2006_3110,bowman2015generating,lopezpaz2016revisiting,danihelka2017comparison,rosca2017variational,im2018quantitatively,gulrajani2020gan,grover2017flowgan,grover2019bias,azadi2018discriminator,turner2018metropolishastings,pmlr-v80-tao18b}.  The most closely related application to this paper is Ref.~\cite{journals/corr/abs-1910-12008} which used learned weights during the training of a generative model to improve fairness.   In this work, the weights are derived after the generator is trained.  In high energy physics, there are many studies using deep generative models~\cite{Paganini:2017hrr,Paganini:2017dwg,Vallecorsa:2019ked,Chekalina:2018hxi,ATL-SOFT-PUB-2018-001,Carminati:2018khv,Vallecorsa:2018zco,Musella:2018rdi,Erdmann:2018kuh,Erdmann:2018jxd,Oliveira:DLPS2017,deOliveira:2017rwa,Hooberman:DLPS2017,Belayneh:2019vyx,Buhmann:2020pmy,deOliveira:2017pjk,Butter:2019eyo,Martinez:2019jlu,Bellagente:2019uyp,Vallecorsa:2019ked,SHiP:2019gcl,Carrazza:2019cnt,Butter:2019cae,Lin:2019htn,DiSipio:2019imz,Hashemi:2019fkn,Zhou:2018ill,Datta:2018mwd,Deja:2019vcv,Derkach:2019qfk,Erbin:2018csv,Urban:2018tqv,Farrell:2019fsm}
and using likelihood ratio estimates based on classifiers~\cite{Andreassen:2019nnm,Badiali:2020wal,Stoye:2018ovl,Hollingsworth:2020kjg,Brehmer:2018kdj,Brehmer:2018eca,Brehmer:2019xox,Brehmer:2018hga,Cranmer:2015bka,Badiali:2020wal,Andreassen:2020nkr,Andreassen:2019cjw,Fischer-ACAT2019}; this paper combines both concepts to construct \textsc{DctrGan}, which is a tool with broad applicability in high energy physics and beyond.

This paper is organized as follows.  Section~\ref{sec:methods} introduces the statistical methods of reweighting and how they can be applied to generative models using deep learning.  Numerical results are presented in Sec.~\ref{sec:results} using standard multimodal probability densities as well as calorimeter simulations from high energy physics.  The paper ends with conclusions and outlook in Sec.~\ref{sec:conclusions}.

\section{Methods}
\label{sec:methods}

\subsection{Generative Models}

A generative model $G:\mathbb{R}^M\rightarrow\mathbb{R}^N$ is a function that maps a latent space (noise) $Z\in\mathbb{R}^M$ to a target feature space $X\in\mathbb{R}^N$.  The goal of the generator is for the learned probability density $p_G(x)$ to match the density of a target process $T$, $p_T(x)$.  There are a variety of approaches for constructing $G$ using flexible parameterizations such as deep neural networks.  Some of these methods produce an explicit density estimation for $p_G$ and others only allow generation without a corresponding explicit density.  

While the method presented in this paper will work for any generative model $G$, our examples will consider the case when $G$ is a Generative Adversarial Network~\cite{Goodfellow:2014upx}.  A \textsc{Gan} is trained using two neural networks:
\begin{align}
\nonumber
(G,D)&=\argmin_{G'}\argmax_{D'} \Big(\mathbb{E}_X[\log D'(X)]\\\label{eq:gan}
&\hspace{25mm}+\mathbb{E}_Z[\log(1-D'(G'(Z))]\Big),
\end{align}
where $D$ is a discriminator/classifier network that distinguishes real examples from those drawn from the generator $G$.  The discriminator network provides feedback to the generator through gradient updates and when it is unable to classify examples as drawn from the generator or from the true probability density, then $G$ will be accurate.  \textsc{Gan}s provide efficient models for generating examples from $G$, but do not generally provide an explicit estimate of the probability density.

This paper describes a method for refining the precision of generative models using a post-processing step that is also based on deep learning using \textit{reweighting}.   A weighting function $W:\mathbb{R}^N\rightarrow [0,\infty)$ is a map designed so that $p_G(x)W(x)\approx p_T(x)$.  Augmenting a generator with a reweighting function will not change the fidelity of any particular example, but it will improve the relative density of examples.  Therefore, this method is most useful for applications that use generators for estimating the ensemble properties of a generative process and not the fidelity of any single example.  For example, generators trained to replace or augment slow physics-based simulators may benefit from the addition of a reweighting function.  Such generators have been proposed for a variety of tasks in high energy physics and cosmology~\cite{Paganini:2017hrr,Paganini:2017dwg,Vallecorsa:2019ked,Chekalina:2018hxi,ATL-SOFT-PUB-2018-001,Carminati:2018khv,Vallecorsa:2018zco,Musella:2018rdi,Erdmann:2018kuh,Erdmann:2018jxd,Oliveira:DLPS2017,deOliveira:2017rwa,Hooberman:DLPS2017,Belayneh:2019vyx,Buhmann:2020pmy,deOliveira:2017pjk,Butter:2019eyo,Martinez:2019jlu,Bellagente:2019uyp,Vallecorsa:2019ked,SHiP:2019gcl,Carrazza:2019cnt,Butter:2019cae,Lin:2019htn,DiSipio:2019imz,Hashemi:2019fkn,Zhou:2018ill,Datta:2018mwd,Deja:2019vcv,Derkach:2019qfk,Erbin:2018csv,Urban:2018tqv,Farrell:2019fsm}.

\subsection{Statistical Properties of Weighted Examples}
\label{sec:weights}

Let $G$ be a trained generative model that is designed to mimic the target process $T$.  Furthermore, let $W(X)$ be a random variable that corresponds to weights for each value of $X$.  If $f$ is a function of the phase space $X$, one can compute the weighted expectation value
\begin{align}
\label{eq:expectation value}
    \mathbb{E}[f(X),W(X)]&=\int p(x)\, W(x)\, f(x) \,dx\,.
\end{align}
Nearly every use of Monte Carlo event generators in high energy physics can be cast in the form Eq.~\ref{eq:expectation value}.  A common use-case is the estimation of the bin counts of a histogram, in which case $f$ is an indicator function that is one if $x$ is in the bin and zero otherwise.  For the true expectation value, $p(x)\mapsto p_T(x)$ and $W(x)\mapsto 1$~\footnote{For simplicity we assume initial weights to be unity. This
procedure can trivially be extended to more complex weight distributions e.g. from
higher-order simulations.}, while for the nominal generator approximation, $p(x)\mapsto p_G(x)$ with $W(x)$ still unity.  The goal of \textsc{DctrGan} is to learn a weighting function $W$ that reduces the difference between $\mathbb{E}_{T}[f(X)]$ and $\mathbb{E}_G[f(X),W(X)]$ for all functions $f$.  The ideal function that achieves this goal is
\beq\label{eq:Wideal}
W(x)={p_T(x)\over p_G(x)}\, .
\eeq

Before discussing strategies for learning $W$, it is important to examine the sampling properties of weighted examples. To simplify the discussion in the remainder of this subsection, we will assume that the ideal reweighting function $W$ has been learned exactly, so that 
\beq
\bar f = \mathbb{E}_{T}[f(X)] = \mathbb{E}_G[f(X),W(X)]\,.
\eeq
for every observable $f$. In practice, small deviations from the ideal reweighting function will result in subleading contributions to the statistical uncertainty\footnote{A related question is the statistical power of the examples generated from $G$.  See Ref.~\cite{Matchev:2020tbw} and~\cite{2008.06545} for discussions, and the latter paper for an empirical demonstration of this topic.}.

Now suppose we generate events $x_i^G$, $i=1,\dots,N_G$, and there are truth events $x_i^T$, $i=1,\dots,N_T$. We are interested in how large $N_G$ must be in order to achieve the same statistical precision as the truth sample. The key observation we will make is that the required $N_G$ depends jointly on the observable $f$ and the weights $W$.

Since we will be interested in variances, it makes sense to consider the mean-subtracted observable
\beq
\delta f = f-\bar f\,.
\eeq
The sampling estimates for $\mathbb{E}_{T}[\delta f(X)]$ and $\mathbb{E}_G[\delta f(X),W(X)]$ are
\bea
& \mathbb{\hat E}_{T}[\delta f(X)] = {1\over N_T}\sum_{i=1}^{N_T} \delta f(x_i^T)\\
& \mathbb{\hat E}_{G}[\delta f(X),W(X)] = {1\over N_G}\sum_{i=1}^{N_G} W(x_i^G) \delta f(x_i^G)\,,
\eea
where hats denote sampling estimates of expectation values. The variances of the sampling estimates are given by\footnote{We are neglecting the contribution to the variance from $\bar f$ which should also be estimated from the sample. It is suppressed by $1/N$ compared to the variance of the expectation value of the sampling estimate of $f$.}
\bea\label{eq:varT}
\text{Var}_T[\mathbb{\hat E}_{T}[\delta f(X)]] &= {1\over N_T} \text{Var}_T[\delta f(X)]\\
&={1\over N_T}\int dx\,p_T(x)\delta f(x)^2\,,
\eea
and
\bea\label{eq:varG}
& \text{Var}_G[\mathbb{\hat E}_{G}[\delta f(X),W(X)]] = {1\over N_G} \text{Var}_G[W(X)\delta f(X)]\\
&\qquad\qquad\qquad ={1\over N_G}\int dx\,p_G(x)W(x)^2\delta f(x)^2\\
&\qquad\qquad\qquad  ={1\over N_G} \int dx\,p_T(x)W(x)\delta f(x)^2\,.
\eea
Equating Eq.~\ref{eq:varT} and~\ref{eq:varG}, we see that to achieve the same statistical precision as the truth events for a given observable $f$, we need to generate 
\beq\label{eq:NGvsNT}
N_G = N_T\times {\int dx\, p_T(x)W(x)\delta f(x)^2\over 
\int dx\, p_T(x)\delta f(x)^2}\,
\eeq
number of events. How many events we need depends on the observable $f$ and the weights $W$. If $W=1$ everywhere (the generator exactly learns $T$), then $N_G=N_T$ for every observable, as expected. Otherwise, if the weights are small (large) in the parts of phase space preferred by $\delta f$, then we will need less (more) events than $N_T$. Smaller weights are better for statistical power -- they correspond to regions of phase space which are over-represented by the generator.

One cannot simply reduce the error everywhere by making all of the weights small.  If the weights are small somewhere in phase space, they must be large somewhere else.  To see this, observe that with the ideal reweighting function, in the large $N$ limit, the weights are automatically normalized across the entire sample such that they sum to $N_G$:
\bea
\sum_{i=1}^{N_G} W(x_i^G) &\to N_G\int dx\,p_G(x)W(x)=N_G\,, \\
\eea
where in the last equation we have substituted Eq.~\ref{eq:Wideal}. So indeed the weights cannot be small everywhere in phase space. 

Evidently, if we want the generator to have good statistical precision across a wide range of observables, we want the weights to be close to unity. Otherwise, there will always be some observables for which we need to generate many more events than $N_T$. 

As a special case of Eq.~\ref{eq:NGvsNT}, consider that $f(x)$ is a histogram bin function, specifically that it is one in a neighborhood of $x=x_0$ and zero otherwise. For a sufficiently small neighborhood, we can ignore the mean $\bar f$ (it is proportional to the volume of the neighborhood), and (\ref{eq:NGvsNT}) reduces to
\beq
N_G = N_T W(x_0)\,.
\eeq
In other words, the value of the weight in the histogram bin directly determines how many events we need to generate to achieve the same statistical precision as the truth sample. 

Finally, while the integral formulas above are formally correct and lead to various insights, they are of limited practical usefulness since we generally do not know $p_T(x)$ and $p_G(x)$ and cannot evaluate the integrals over $x$. For actually estimating the uncertainties on expected values, one should replace the integrals in Eq.~\ref{eq:varT} and Eq.~\ref{eq:varG} with their sampling estimates.

\subsection{Learning the Weighting Function}
\label{sec:weighting}

The weighting function $W$ is constructed using the \textit{Deep neural networks using Classification for Tuning and Reweighting} (\textsc{Dctr}) methodology~\cite{Andreassen:2019nnm} (see also Ref.~\cite{Badiali:2020wal,Stoye:2018ovl,Hollingsworth:2020kjg,Brehmer:2018kdj,Brehmer:2018eca,Brehmer:2019xox,Brehmer:2018hga,Cranmer:2015bka,Badiali:2020wal,Andreassen:2020nkr,Andreassen:2019cjw}).  Let $\mathcal{X}_\text{real}$ be a set of examples drawn from $p_X(x)$ and $\mathcal{X}_\text{fake}$ be a set of examples drawn from $p_Z(z)$ and then mapped through $G$, henceforth called $p_G(x)$ for $x=G(z)$.  A neural network classifier $h:\mathbb{R}^N\rightarrow [0,1]$ is trained to distinguish $\mathcal{X}_\text{real}$ from $\mathcal{X}_\text{fake}$ using the binary cross entropy loss function
so that $h(x)=1$ corresponds to $\mathcal{X}_\text{real}$
\footnote{A variety of other loss functions can also be applied such as the mean squared error.  The dependence of $W$ on $h$ may need to be modified if other loss functions are used.}.  The weighting function is then constructed as $W(x)=h(x)/(1-h(x))$.  With sufficient training data, classifier expressiveness, and training flexibility, one can show that $W$ approaches the likelihood ratio $p_T(x)/p_G(x)$.  If the weights are exactly the likelihood ratio, then the expectation values computed in Eq.~\ref{eq:expectation value} will be unbiased.

Training a classifier is often easier and more reliable than training a full generative model.  Therefore, a natural question is why learn $G$ at all?  The answer is because the further $W(x)$ is from the constant unity function, the more statistical dilution will arise from the generated examples (when a sufficiently broad range of observables is considered), as described in Sec.~\ref{sec:weights}.
In particular, this method does not work at all when there are weights that are infinite, corresponding to regions of phase space where $p_T(x)>0$ but $p_G(x)=0$.  Therefore, the \textsc{Dctr} approach should be viewed as a refinement and the better $G$ is to begin with, the more effective $W$ will be in refining the density.
Our method differs from the refinement proposed in Ref.~\cite{Erdmann:2018kuh} because \textsc{Dctr}-ing leaves the data-points intact and only changes their statistical 
weight while~\cite{Erdmann:2018kuh} modifies the data-points themselves. 
A combined framework that simultaneously changes both data and weights might be 
an interesting avenue for future studies.

\section{Results}
\label{sec:results}

The \textsc{DctrGan} approach is illustrated with two sets of examples: one set where the true density $p_T$ is known analytically (Sec.~\ref{sec:multimodal}) and one where $p_T$ is not known, but samples can be drawn from it (Sec.~\ref{sec:calosim}).  

\subsection{Multimodel Gaussian Distributions}
\label{sec:multimodal}

As a first set of examples, Gaussian mixture models are simulated in one and two dimensions.   Multimodal probability densities are traditionally challenging for standard \textsc{Gan}s to model precisely and are a useful benchmark for the \textsc{DctrGan} methodology.   In particular, the following random variables are simulated:
\begin{enumerate}
\item \textit{1D bimodal}: 
\begin{align}
p_X(x)=\frac{1}{2}N(x,-1,0.4)+\frac{1}{2}N(x,1,0.4),
\end{align}
where $N(x,\mu,\sigma)$ is the probability density of a normal random variable with mean $\mu$ and standard deviation $\sigma$ evaluated at $x$.
\item \textit{2D circular}:
\begin{align}\nonumber
p_X(x)&=\frac{1}{8}\sum_{i=1}^8\left[N\left(x_0,\cos\left(\frac{2\pi i}{8}\right),0.1\right)\right.\\
&\hspace{15mm}\times \left.N\left(x_1,\sin\left(\frac{2\pi i}{8}\right),0.1\right)\right],
\end{align}
where $x=(x_0,x_1)\in\mathbb{R}^2$.

\item \textit{2D $5\times 5$ grid}:
\begin{align}\nonumber
p_X(x)&=\frac{1}{25}\sum_{i=1}^5\sum_{j=1}^5\left[N\left(x_0,\left(i+0.1\right),0.1\right)\right.\\
&\hspace{25mm}\times \left.N\left(x_1,j+0.1,0.1\right)\right].
\end{align}

\end{enumerate}

All models were implemented in \textsc{Keras}~\cite{keras} with the \textsc{Tensorflow} backend~\cite{tensorflow} and optimized with the \textsc{Adam}~\cite{adam}.  The discriminator $D$ networks have two hidden layers with 25 nodes each and use the rectified linear unit (ReLU) activation function.  The sigmoid function is used after the last layer.   The generator networks $G$ also has two hidden layers, with 15 units each.  A latent space of dimension 5, 8, and 25 is used for the 1D bimodal, 2D circular, and 2D $5\times 5$ grid, respectively.  Each model was trained with a batch size of 128 for 10,000 epochs (passes through the batches, not the entire dataset).  For the reweighting model, three hidden layers were used for all three cases with ReLU activiations on the intermeidate layers and a sigmoid for the last layer.  For the 2D models, the numbers of hidden units were 64, 128, and 256 while the 1D example used 20 hidden notes on each intermediate layer.  The binary cross entropy was used for the loss function and the models were trained for 100 epochs (passes through the entire dataset).

\begin{figure}[h!]
\centering
\includegraphics[width=0.5\textwidth]{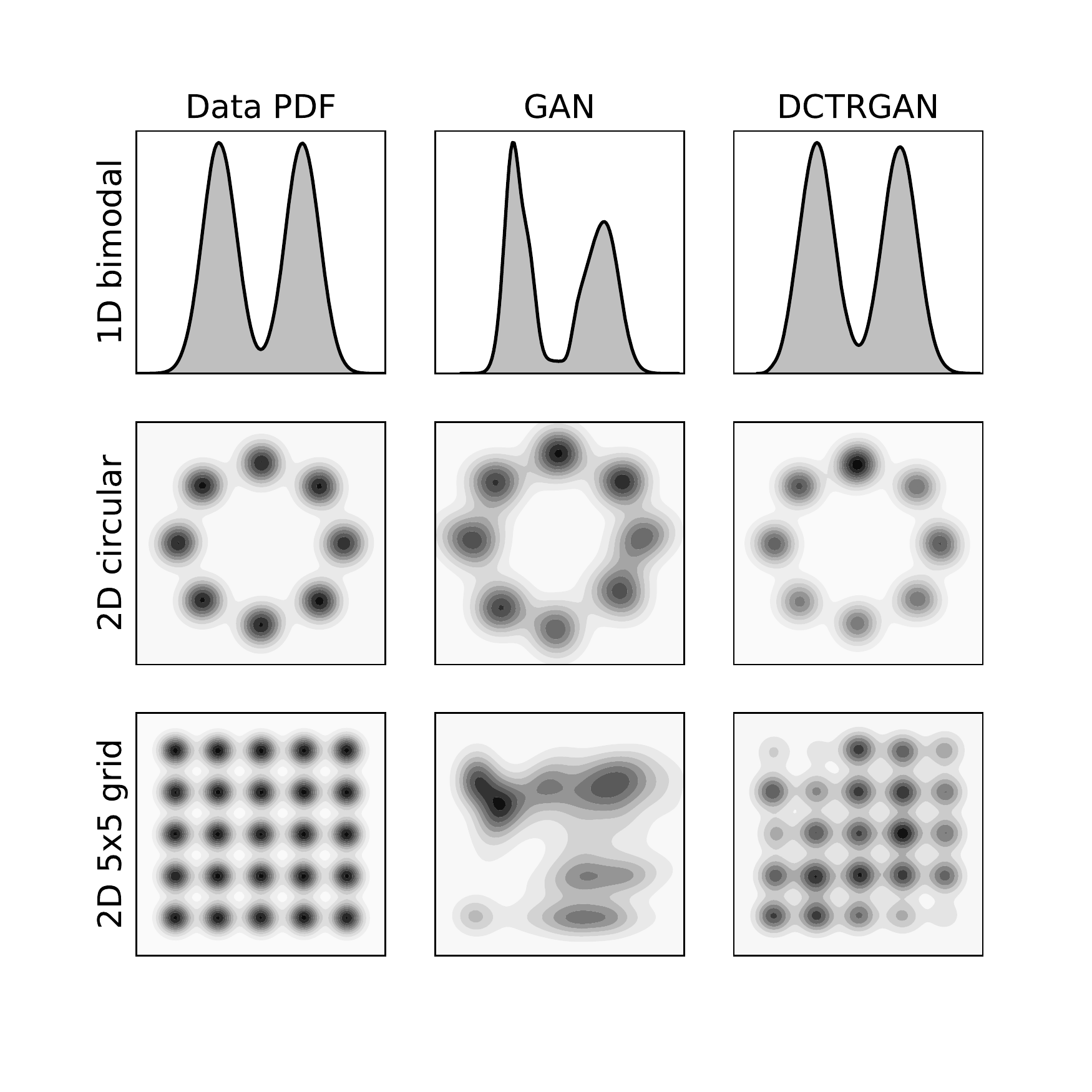}
\caption{Three multi-modal Gaussian examples.  The true probability density $p_X$ is shown in the first column, a \textsc{Gan} density is presented in the second column, and then the density from \textsc{DctrGan} is drawn in the third column.  Examples inspired from Ref.~\cite{fisher2018boltzmann}.}
\label{fig:simple}
\end{figure}

The resulting probability densities are presented in Fig.~\ref{fig:simple}.  In all cases, the fidelity of the \textsc{Gan} density is significantly improved using reweighting. 

\subsection{Calorimeter Simulation Examples}
\label{sec:calosim}

Data analysis in high energy physics makes extensive use of simulations for inferring fundamental properties of nature.  These complex simulations encode physics spanning a vast energy range.  The most computationally demanding part of the simulation stack is the modeling of particles stopping in the dense material of calorimeters that part of most detectors.  \textsc{Gan}s have been investigated as a surrogate model for accelerating these slow calorimeter simulations~\cite{Paganini:2017hrr,Paganini:2017dwg,Vallecorsa:2019ked,Chekalina:2018hxi,ATL-SOFT-PUB-2018-001,Carminati:2018khv,Vallecorsa:2018zco,Musella:2018rdi,Erdmann:2018kuh,Erdmann:2018jxd,Oliveira:DLPS2017,deOliveira:2017rwa,Hooberman:DLPS2017,Belayneh:2019vyx,Buhmann:2020pmy}  

The \textsc{Gan} model studied here is a modified version of the Bounded-Information-Bottleneck autoencoder (BIB-AE)~\cite{voloshynovskiy2019information} shown in Ref.~\cite{Buhmann:2020pmy}. The BIB-AE setup is based on the encoder-decoder structure of a VAE, but its training is enhanced through the use of Wasserstein-\textsc{Gan}~\cite{pmlr-v70-arjovsky17a}-like critics. The theoretical basis of the model is discussed in Ref.~\cite{voloshynovskiy2019information}, while the explicit architecture and training process is described in the appendix of Ref.~\cite{Buhmann:2020pmy}. The main modification to Ref.~\cite{Buhmann:2020pmy} is introduced to reduce \textit{mode collapse}: regions of phase space that are significantly under- or over-sampled from the \textsc{Gan}.  If extreme enough, such regions can cause \textsc{Dctr} weights that lead to infinites in the loss function.  Our modified version maintains the encoder and decoder architecture, but each critic is replaced by a set of two identical critic networks. These two critics are trained in parallel, however one of the two has its weights reset after every epoch. Based on its training history the continuously trained critic may be blind to certain artifacts in the generated shower images that lead to mode collapse. The reset critic, however, is able to notice and reduce such artifacts. Additionally we change the input noise to be uniform instead of Gaussian and skip the neural network based post processing described in Ref.~\cite{Buhmann:2020pmy}, as most of its effects can be replicated through the \textsc{Dctr} approach.

The real examples $T$ are based on detailed detector simulations using \textsc{Geant4} 10.4~\cite{Agostinelli:2002hh} through a DD4hep 1.11~\cite{Frank_2014} implementation of the International Large Detector (ILD)~\cite{ILD:2020qve} proposed for the International Linear Collider.  The calorimeter is composed of 30 active silicon layers with tungsten absorber.  The energy recorded in each layer is projected onto a regular grid of $30\times 30$ cells. To simulate the ILD Minimal Ionizing Particle (MIP) cutoff we removes hits with energies $<0.1~\text{MeV}$ for both the \textsc{Geant4} and \textsc{Gan} showers. All showers are generated for an incident photon with an energy of 50 GeV. More details
on the simulation are given in Ref.~\cite{Buhmann:2020pmy}.

Two different \textsc{Dctr} models are trained: one using the original data dimensionality (low-level or LL) and one that uses slightly processed inputs with lower dimensionality.   The latter network is built from 33 features: the number of non-zero cells, the longitudinal centroid (energy-weighted layer index of the shower: $\sum_{i=1}^{30} i \,E_i/\sum_{i=1}^{30} E_i$), the total energy, and the energy in each of the thirty layers.  A fully connected network processes these observables, with two hidden layers of 128 nodes each. The ReLU activation is used for the intermediate layers and the sigmoid is used for the final layer and the model is trained with the binary cross entropy loss function. The network was implemented using \textsc{Keras} with the \textsc{Tensorflow} backend and optimized with the \textsc{Adam} and optimized with a batch size of 128 for 50 epochs.  

The low-level classifier was trained directly on the $30\times30\times30$ shower images.  On major problem in this approach are artificial features in the \textsc{Gan} showers caused by mode collapse. These give the classifier a way to distinguish between \textsc{Geant4} and \textsc{Gan} without learning the actual physics. This in turn means that the showers with a high classification score are not necessarily the most realistic ones, which reduces the effectiveness of the reweighting. Therefore a large part of the low-level classifier setup is designed to reduce the impact of such artifacts.  The feature processing for the high-level network acts as an effective regularization that mitigates these effects.

The low-level classifier was built using \textsc{PyTorch}~\cite{NEURIPS2019_9015}. The initial $30\times30\times30$ input image first has a 50\% chance of being flipped along the $x$ direction and another 50\% chance of being flipped along the $y$ direction.  These directions are perpendicular to the longitudinal direction of the incoming photon.  The input is largely symmetric under these transformations, but they make it harder for the classifier to pick up on the above mentioned artifacts. The image is then  passed through 3 layers of 3D-convolutions, all with 128 output filters and $3\times3\times3$ size kernels. The first two convolutions have a stride of 2 with a zero padding of 1, while the third has a stride of 1 and no padding. Between the first and second and second and third convolution layer-norm steps are performed. The $128\times6\times6\times6$ output of the final convolutions is flattened and passed to a set of fully connected layers with (64, 256, 256, 1) output nodes, respectively. Each layer uses a \textsc{LeakyReLU} activation function~\cite{Maas13rectifiernonlinearities} with a slope of 0.01, except for the final output layer, which uses a sigmoid activation. The network is trained with a binary cross entropy loss function using the \textsc{Adam} optimizer with a learning rate of $10^{-3}$. The training set consist of 500k \textsc{Gan} showers and 177k \textsc{Geant4} showers, the \textsc{Geant4} set is tiled, so it matches the size of the \textsc{Gan} set.  This is equivalent to increasing the training weights of the \textsc{Geant4} set, but was found to converge better than using weights. The network is trained for 1 epoch, as longer training makes the classifier more sensitive to the artificial structures in the \textsc{Gan} showers. After training the classifier is calibrated using Temperature Scaling~\cite{10.5555/3305381.3305518}. Finally we clip the individual per shower weight to be no larger than 5.

\begin{figure}
\centering
\includegraphics[width=0.4\textwidth]{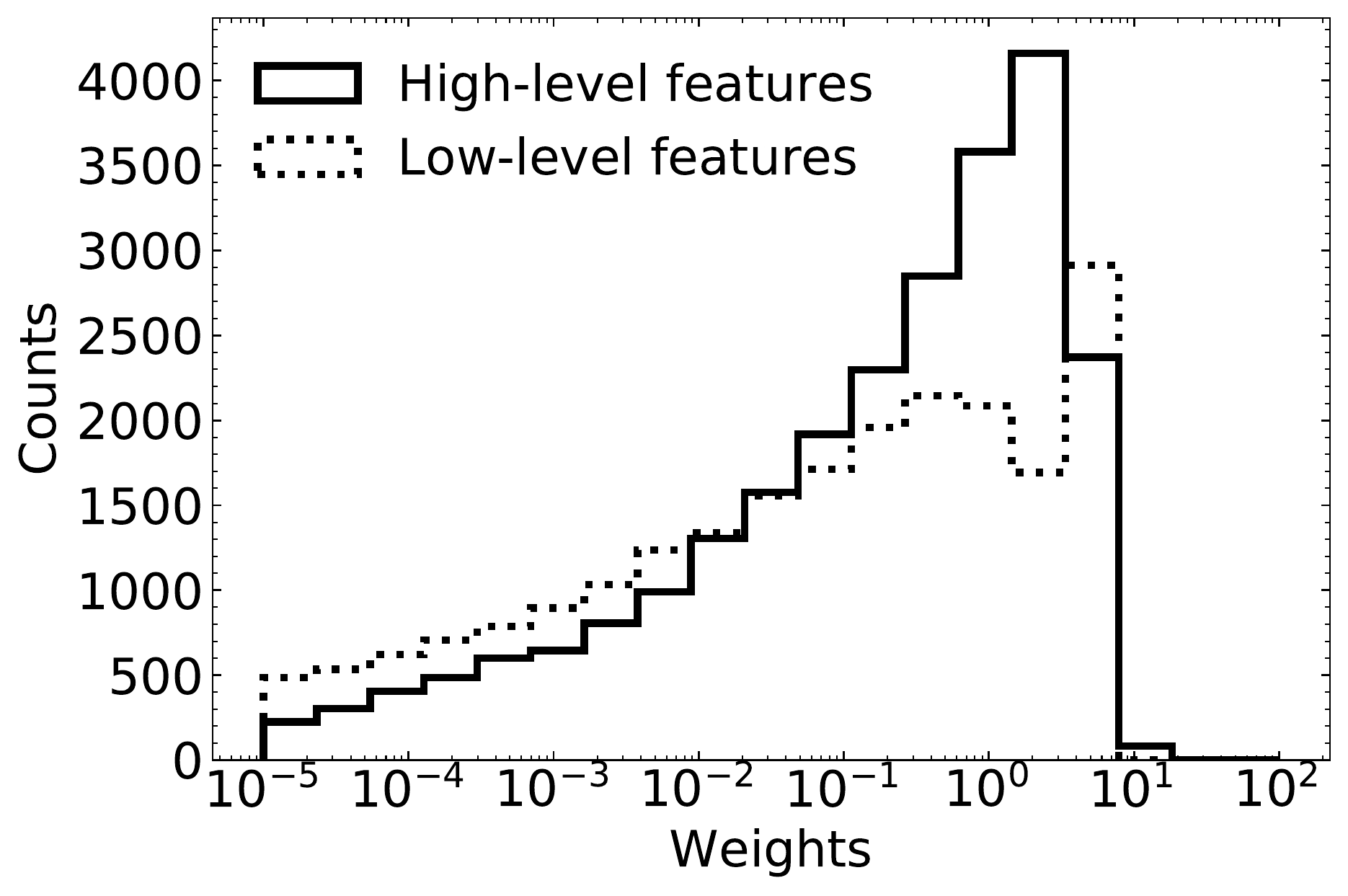}\\\includegraphics[width=0.4\textwidth]{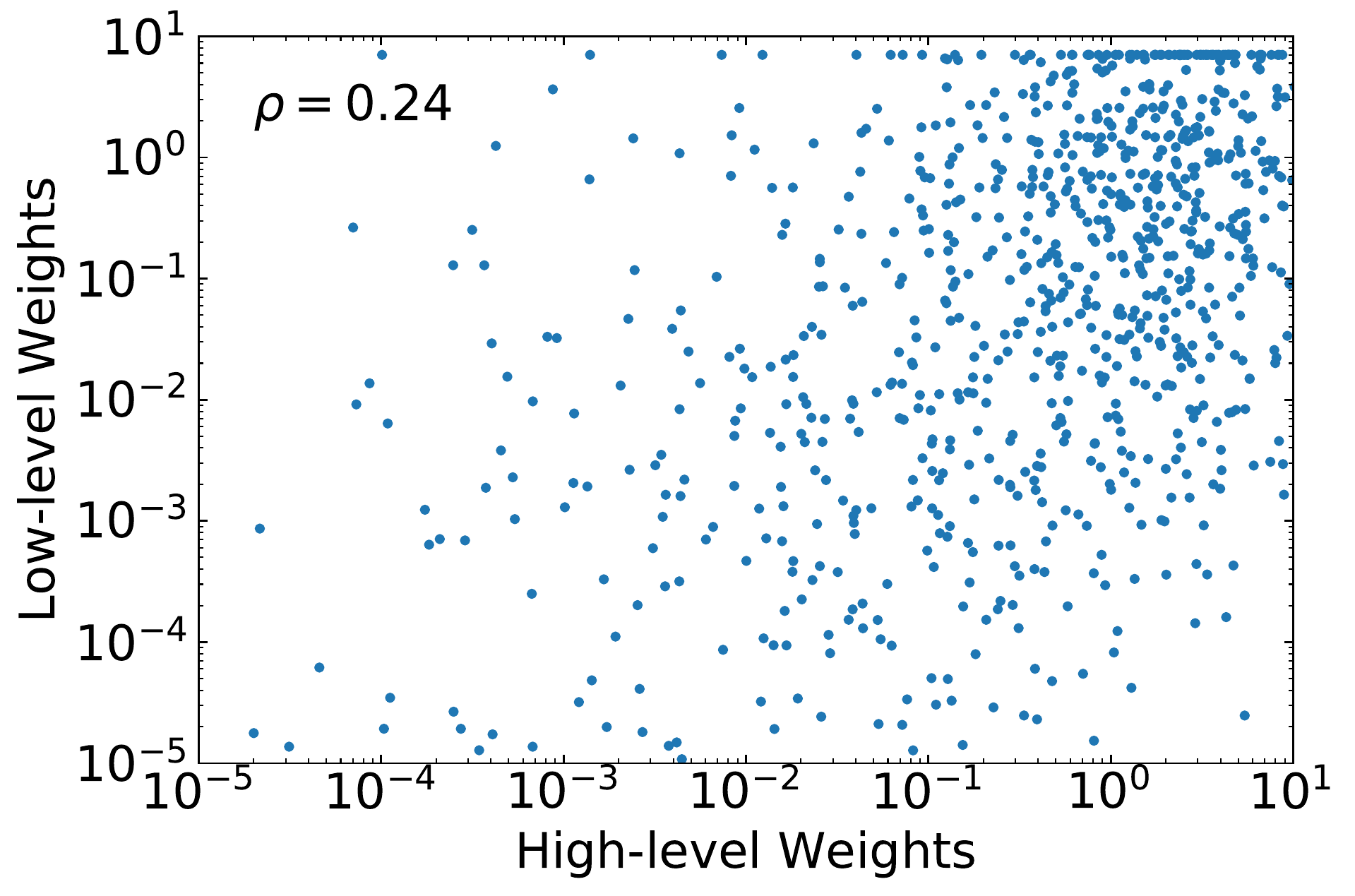}
\caption{The weights of the low-level and high-level \textsc{Dctr} models.  The top plot presents histograms of the weights and the bottom plot presents a scatter plot demonstrating the correlation between the weights of the two models.  The Pearson correlation ($\rho$) is indicated in the plot.}
\label{fig:weights}
\end{figure}

Figure~\ref{fig:weights} shows histograms of the learned \textsc{Dctr} weights for both the low-level and high-level models.  The most probable weight is near unity, with a long and asymmetric tail towards low weights.  The right plot of Fig.~\ref{fig:weights} shows that there is a positive correlation between the weights of the two models.

\begin{figure}
\centering
\includegraphics[width=0.4\textwidth]{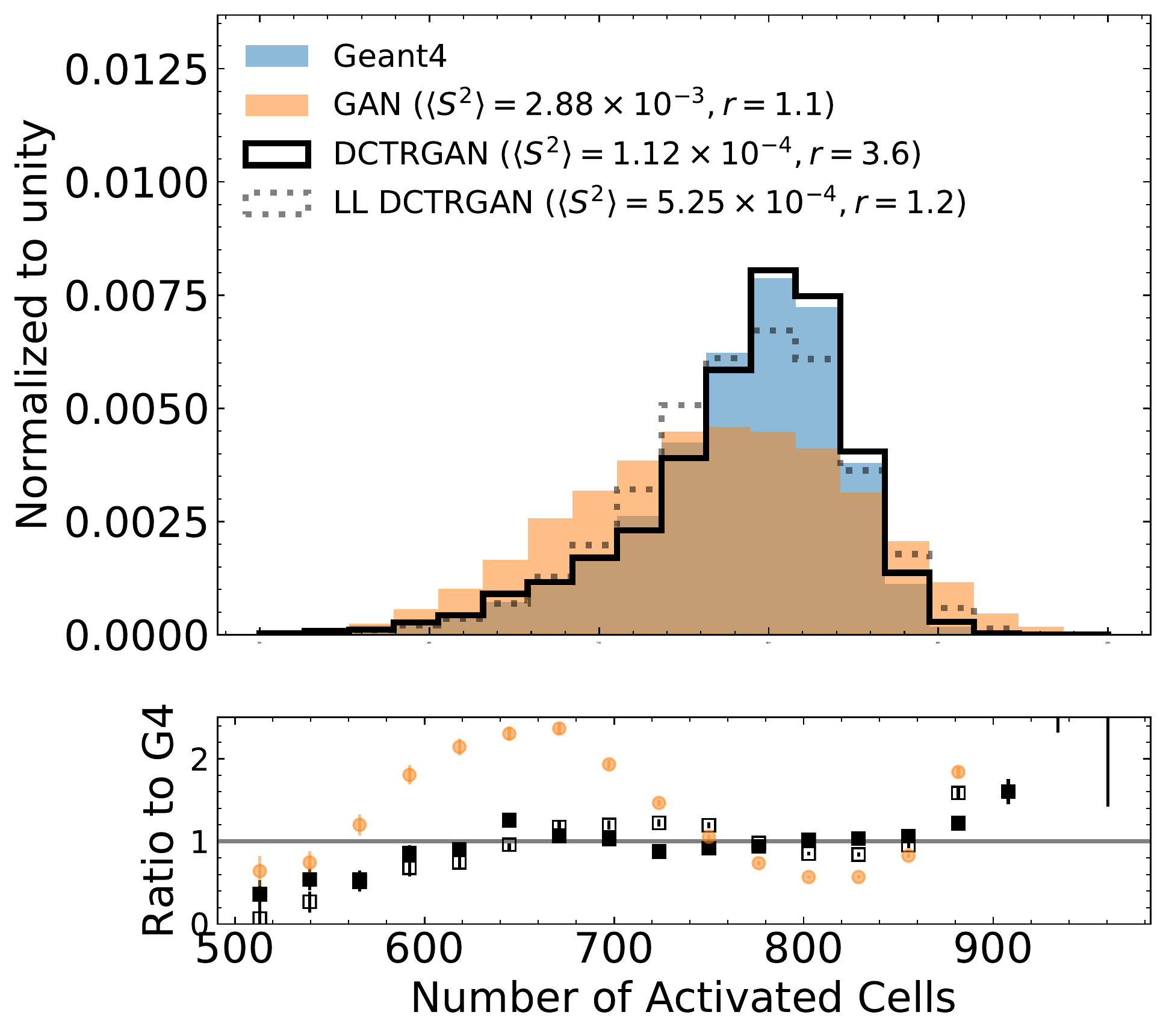}\\\includegraphics[width=0.4\textwidth]{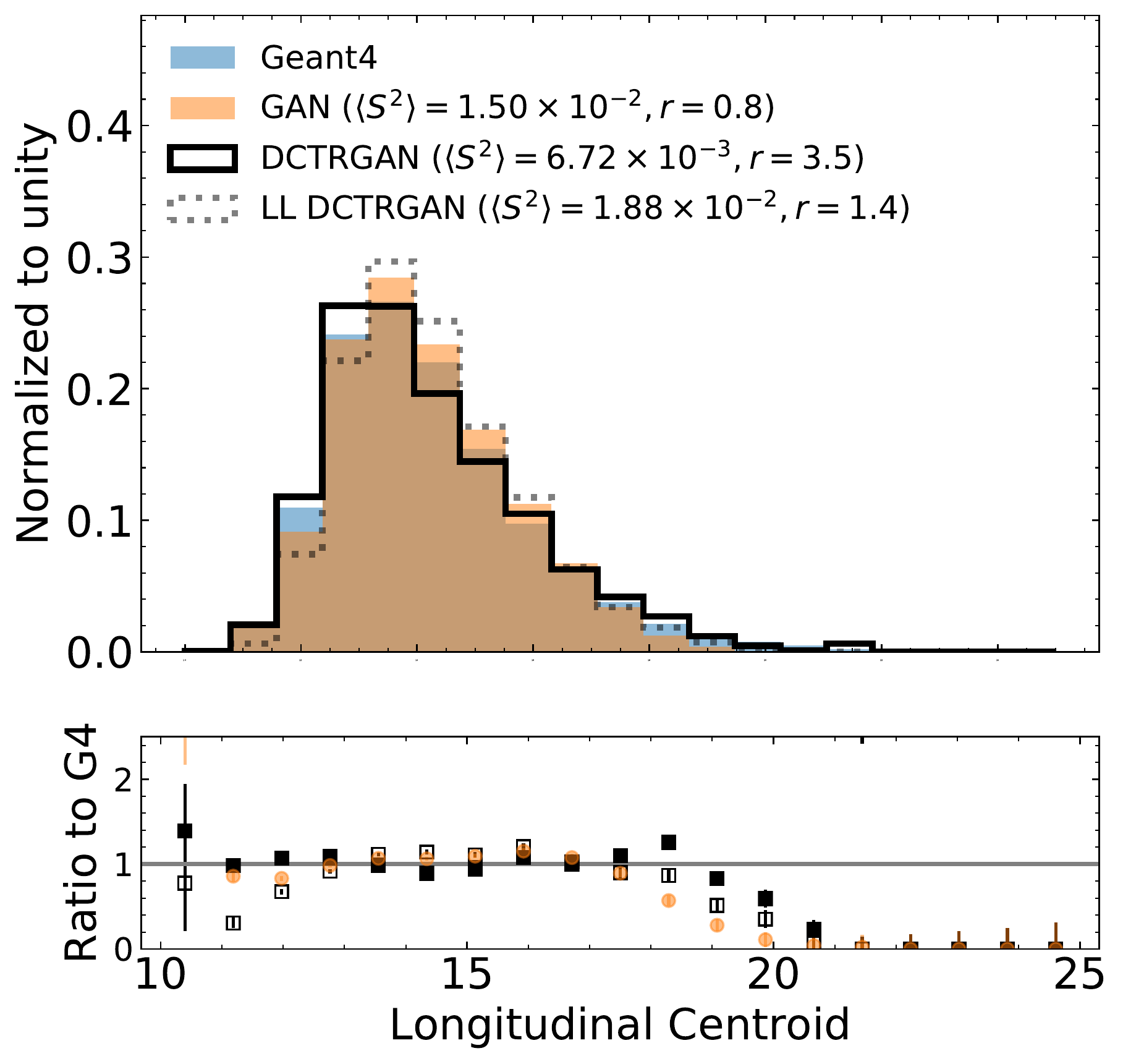}\\\includegraphics[width=0.4\textwidth]{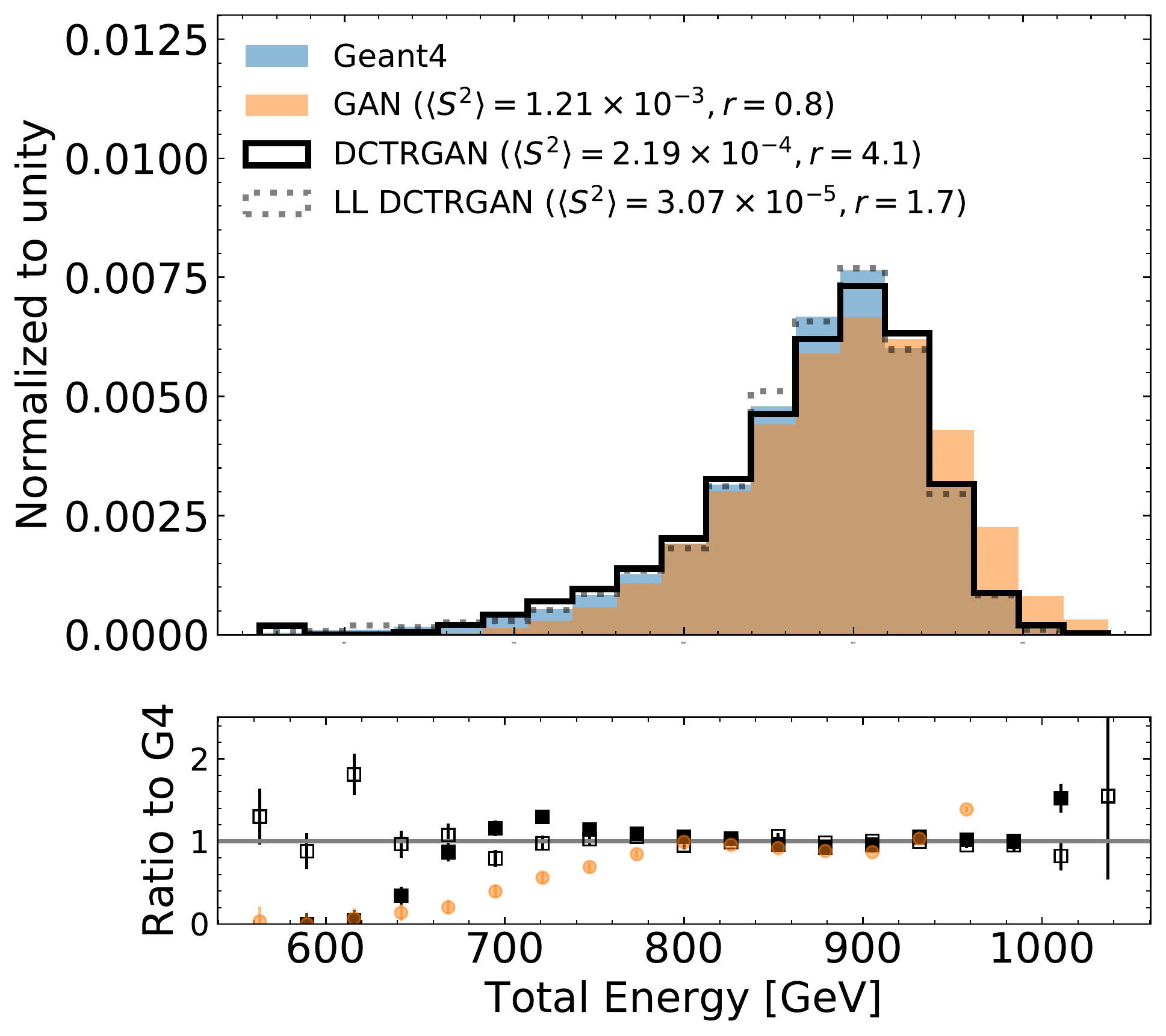}\\
\caption{Histograms of various observables from simulated calorimeter showers of 50 GeV photons in a 5-layer calorimeter with $30\times 30$ cells in each layer.  A cell is activated if a non-zero energy is registered in that cell. The panels below each histogram show the ratio between the \textsc{Gan} or the \textsc{DctrGan} and the physics-based simulator \textsc{Geant4}.  The legend includes the separation power $\langle S^2\rangle$ between the (weighted) \textsc{Gan} model and the \textsc{Geant}4 model.  Additionally, the ratio $r$ of the uncertainty in the mean of the observable between the \textsc{Gan} and \textsc{Geant}4 is also presented. }
\label{fig:caloexamples}
\end{figure}

\begin{figure*}
\centering
\includegraphics[height=0.28\textwidth]{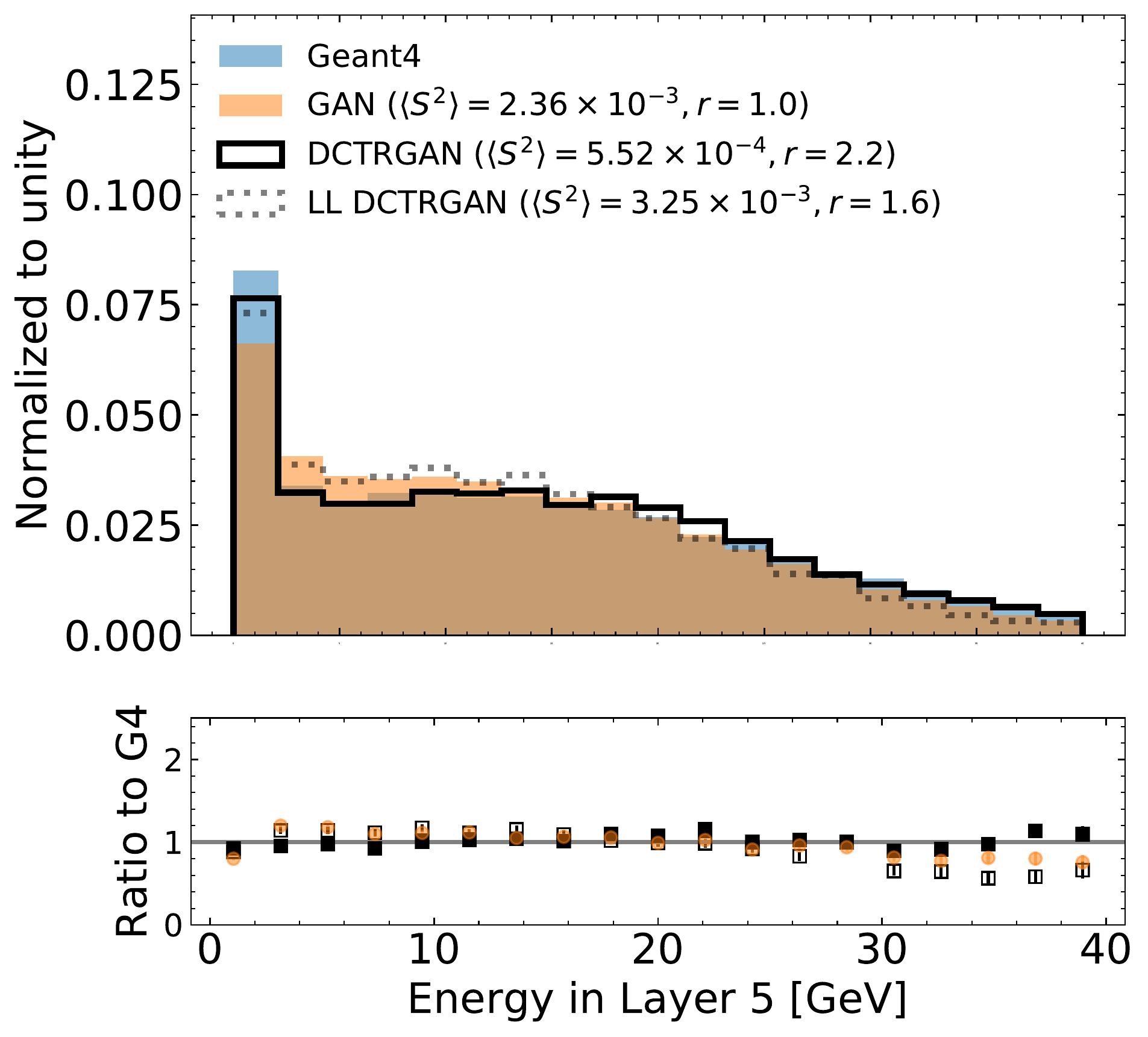}
\includegraphics[height=0.28\textwidth]{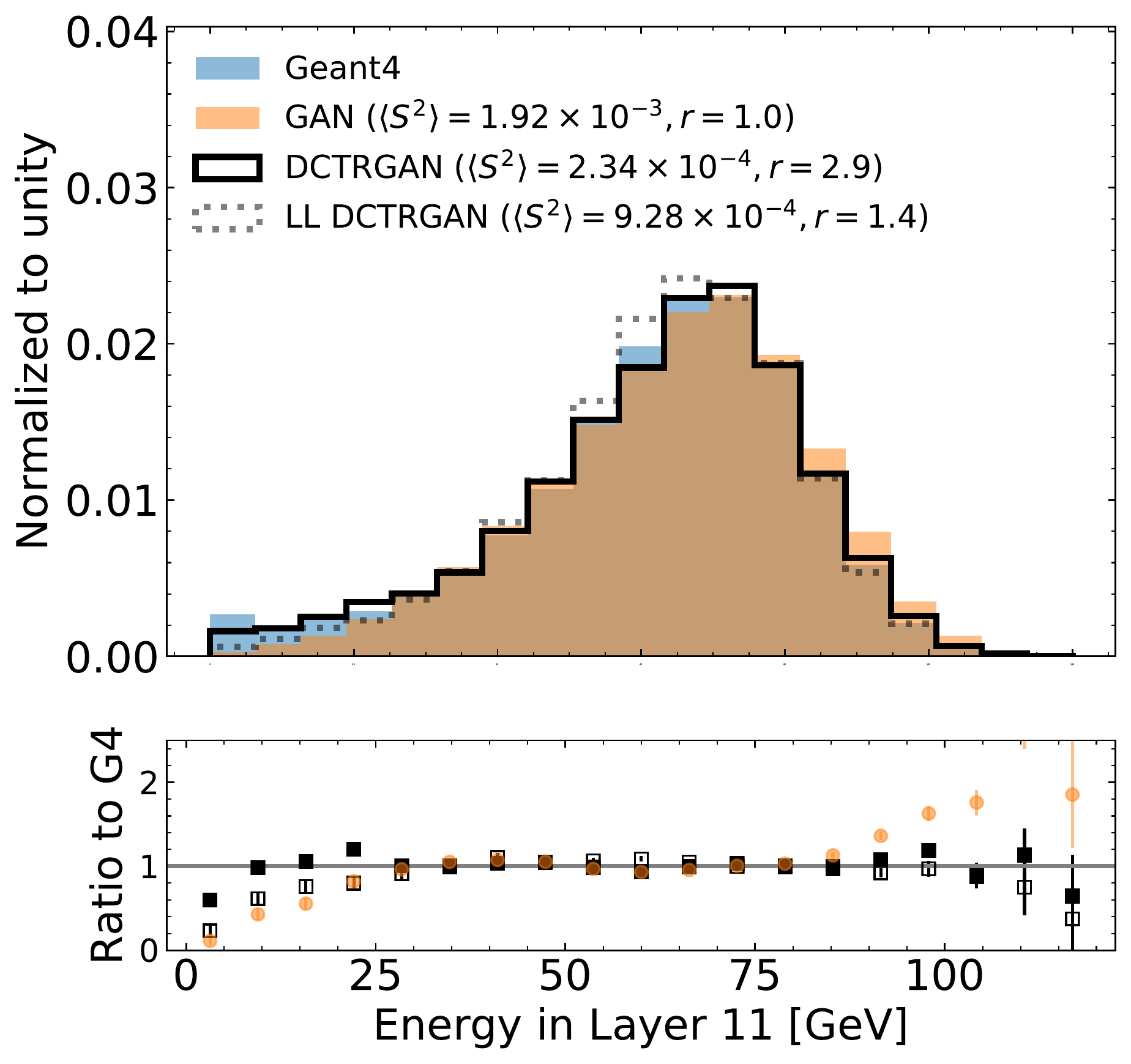}
\includegraphics[height=0.28\textwidth]{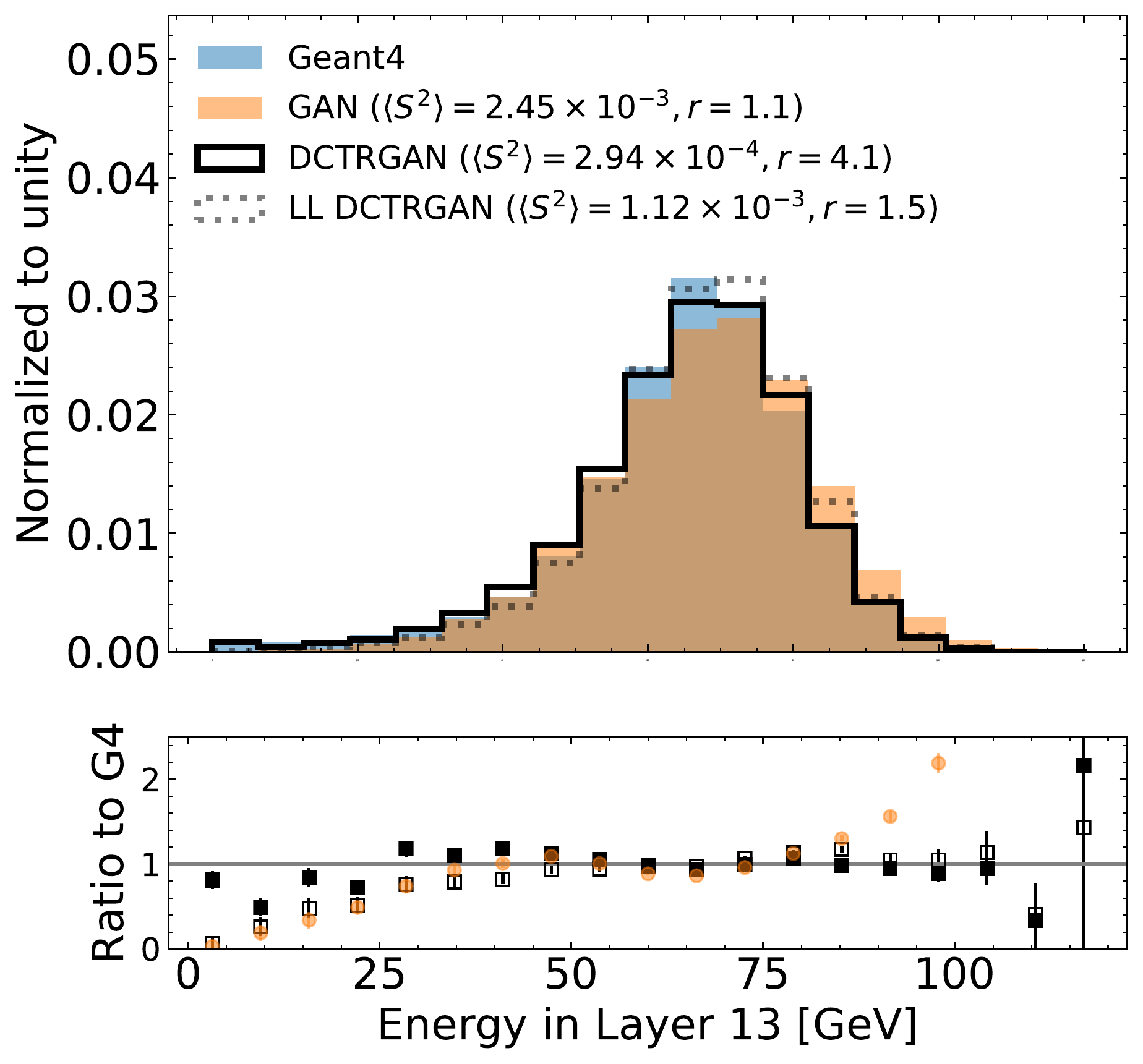}\\
\includegraphics[height=0.28\textwidth]{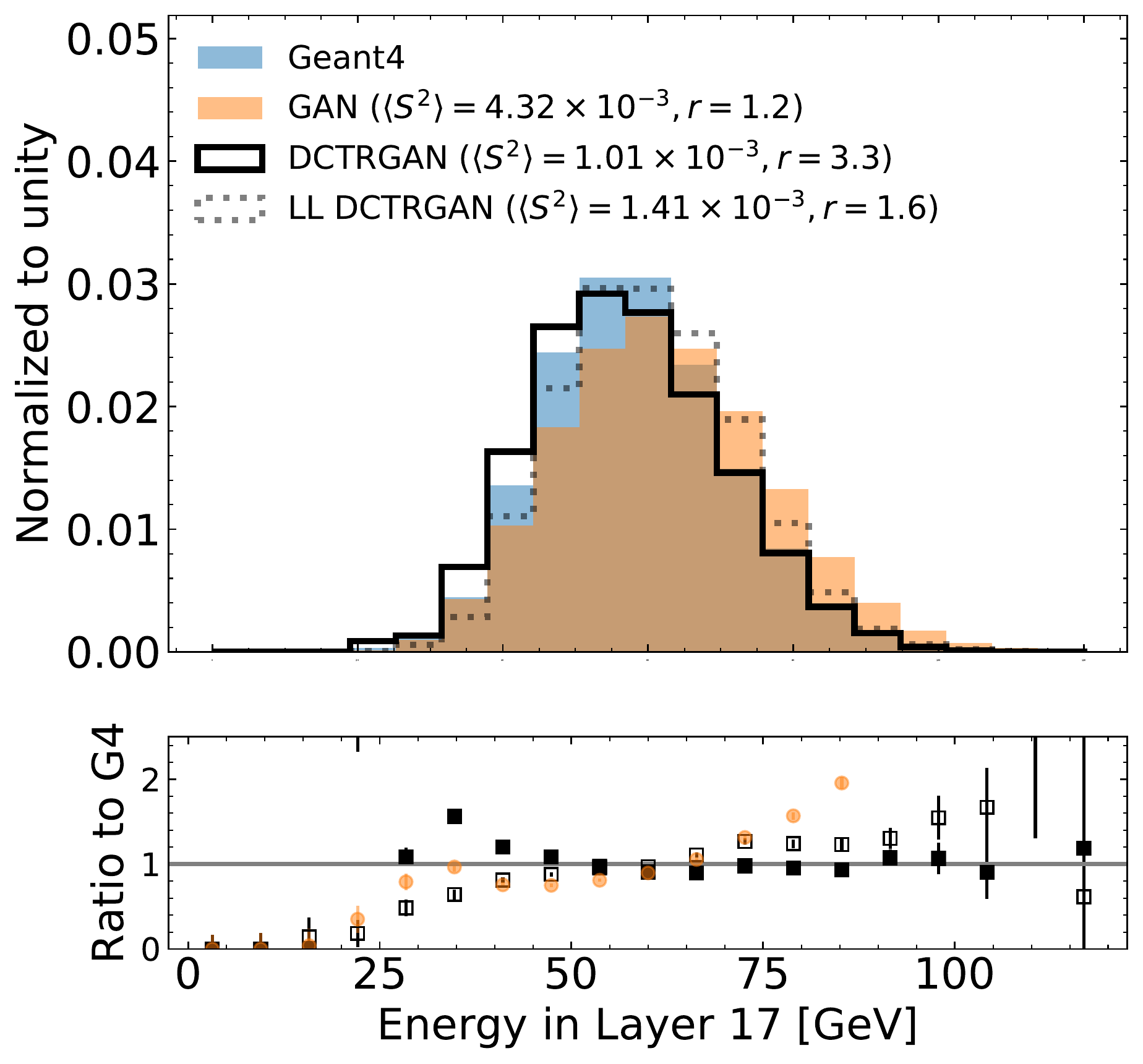}
\includegraphics[height=0.28\textwidth]{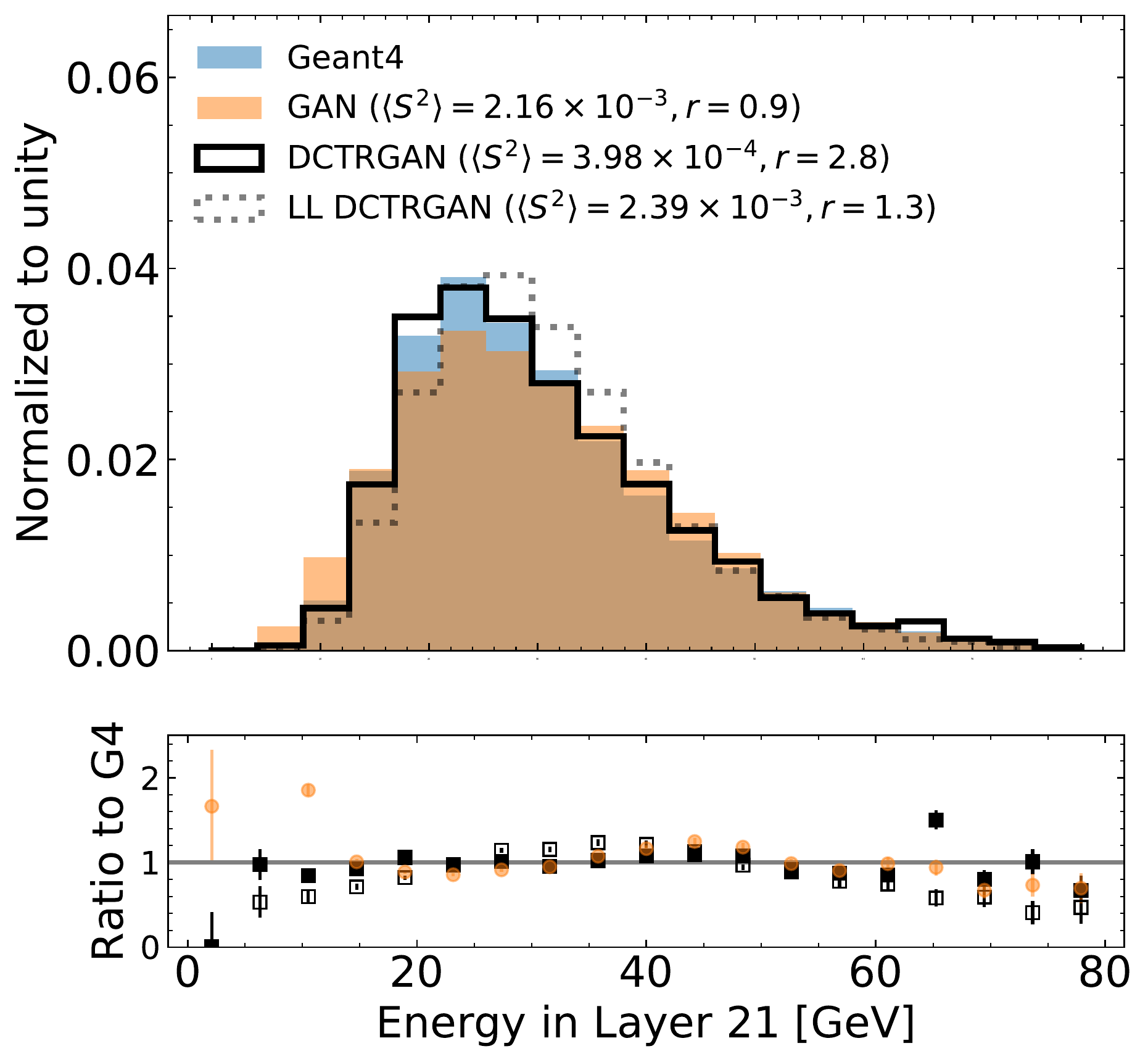}
\includegraphics[height=0.28\textwidth]{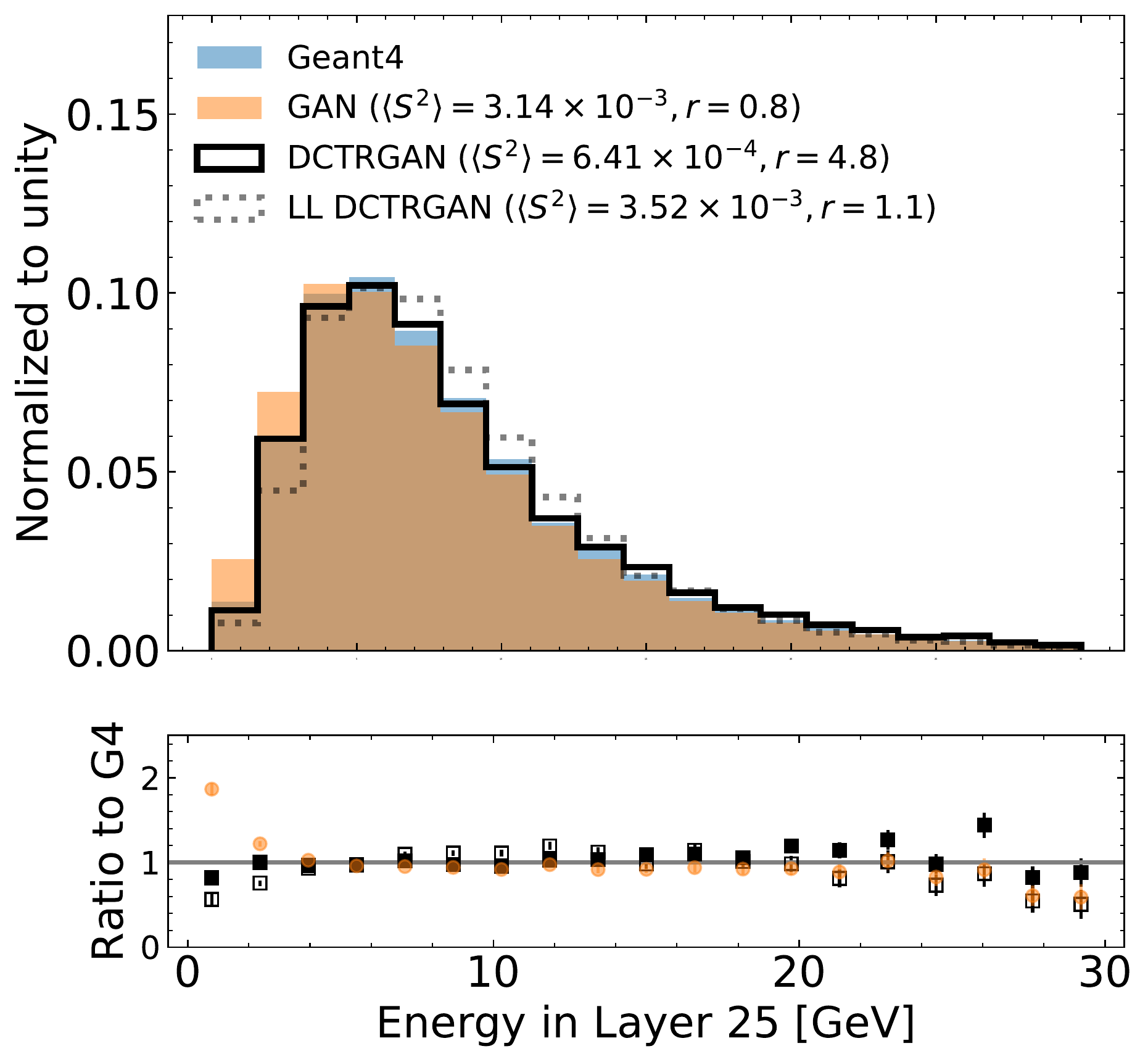}
\caption{Histograms of total energies in selected layers of the calorimeter.  The panels below each histogram show the ratio between the \textsc{Gan} or the \textsc{DctrGan} and the physics-based simulator \textsc{Geant4}.  The legend includes the separation power $\langle S^2\rangle$ between the (weighted) \textsc{Gan} model and the \textsc{Geant}4 model.  Additionally, the ratio $r$ of the uncertainty in the mean of the observable between the \textsc{Gan} and \textsc{Geant}4 is also presented.  Underflow and overflow are not included in the leftmost or rightmost bins.}
\label{fig:caloexamples2}
\end{figure*}

\begin{figure*}
\centering
\includegraphics[height=0.22\textwidth]{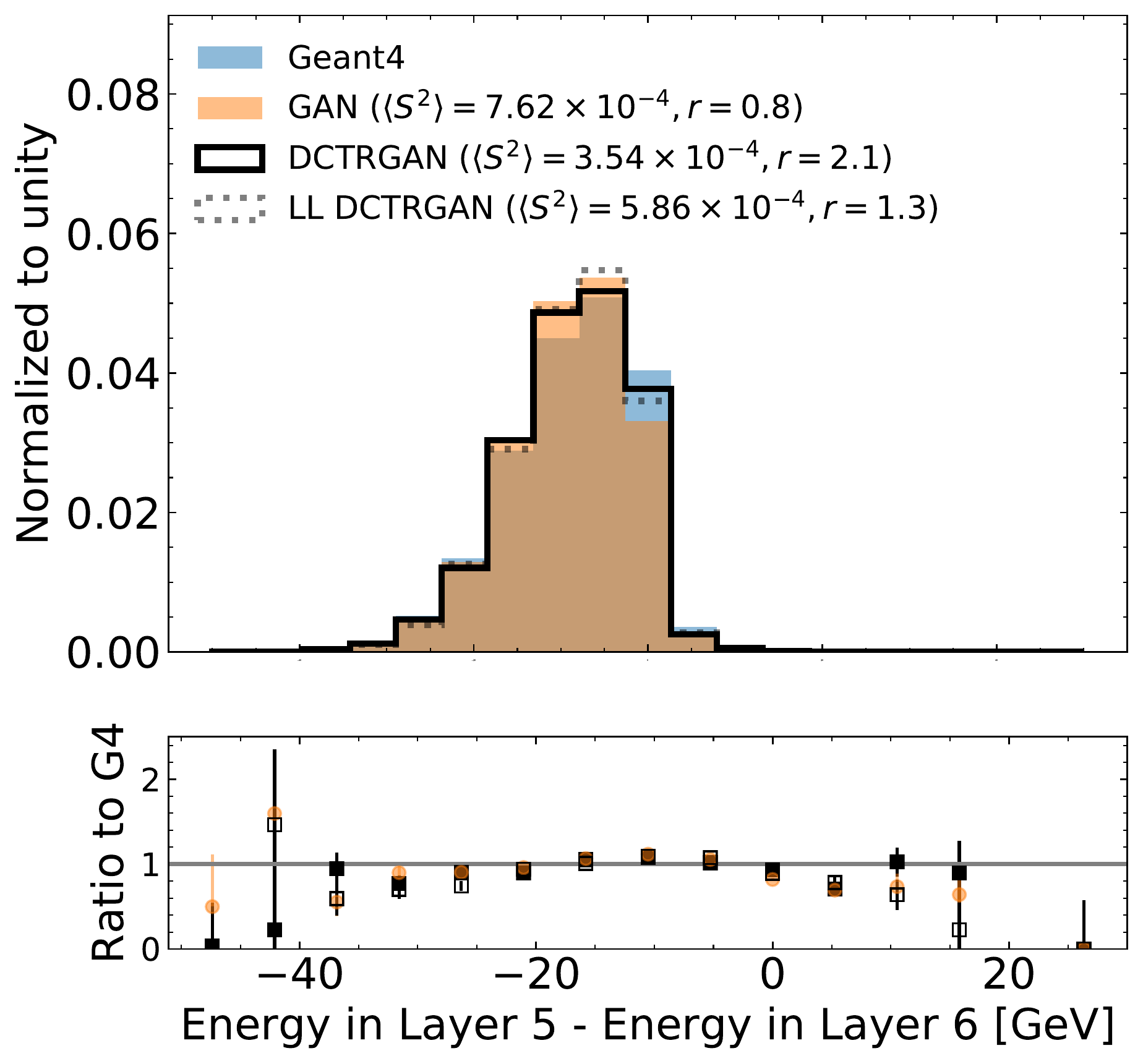}
\includegraphics[height=0.22\textwidth]{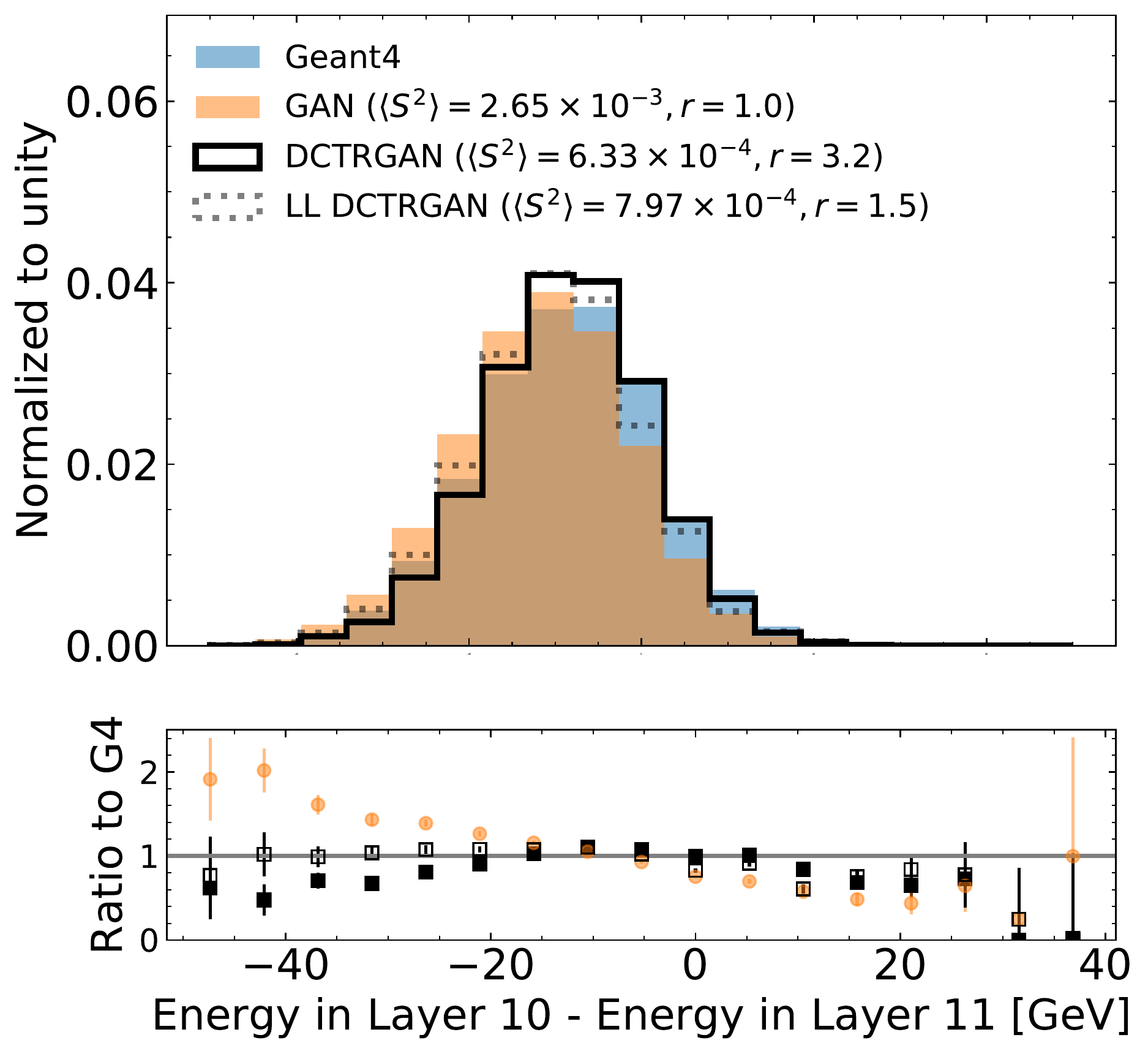}
\includegraphics[height=0.22\textwidth]{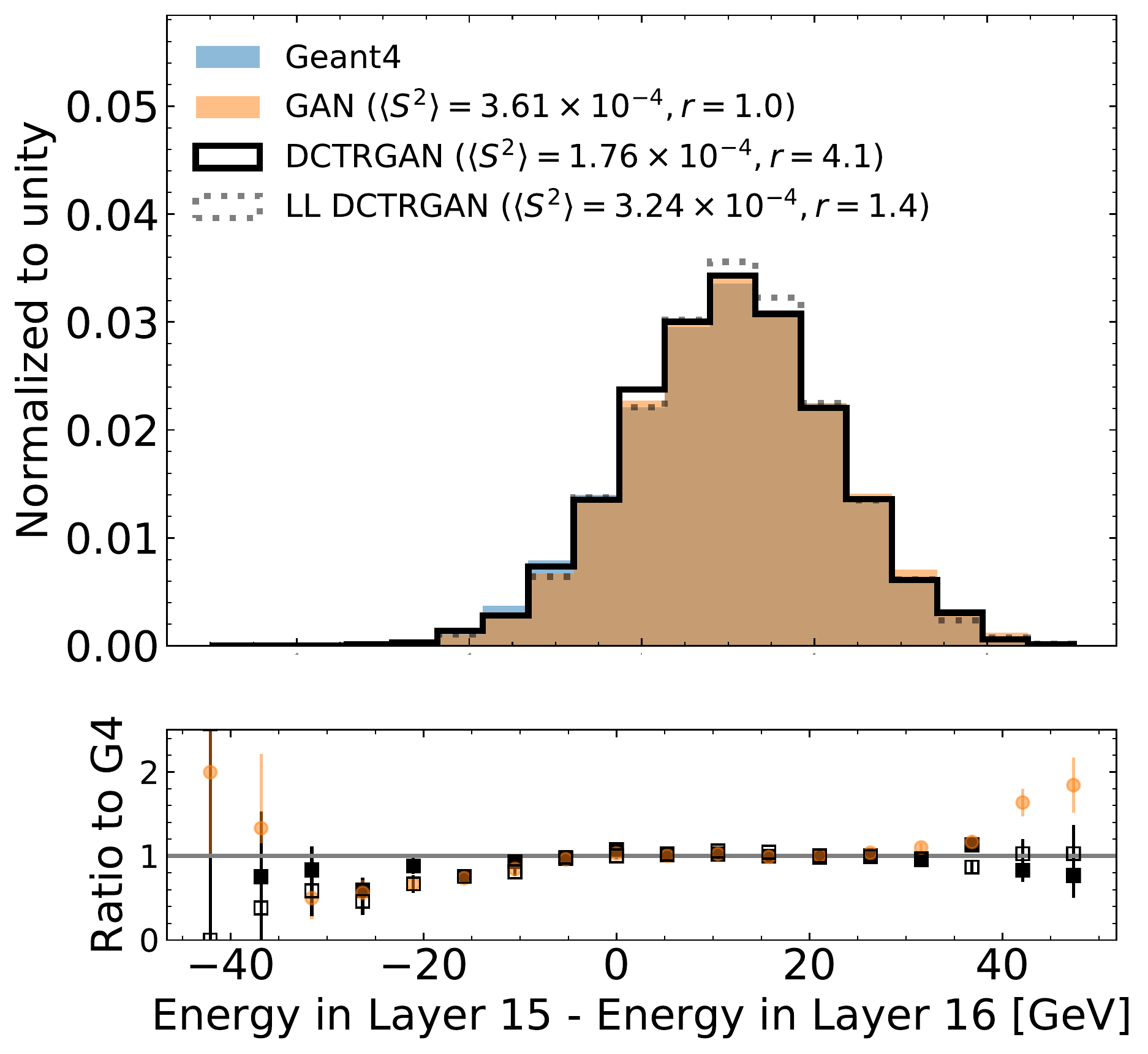}
\includegraphics[height=0.22\textwidth]{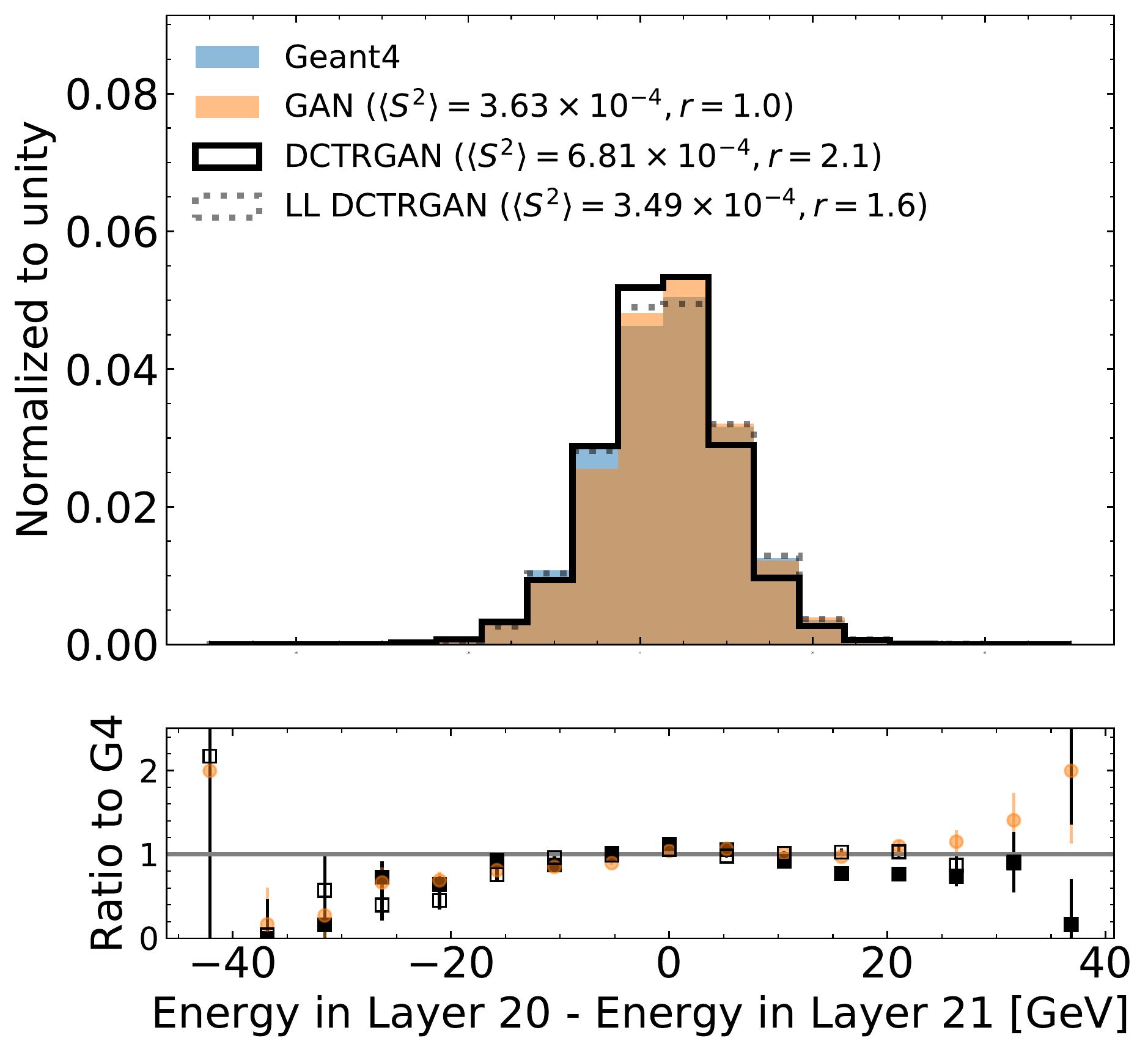}\\
\includegraphics[height=0.22\textwidth]{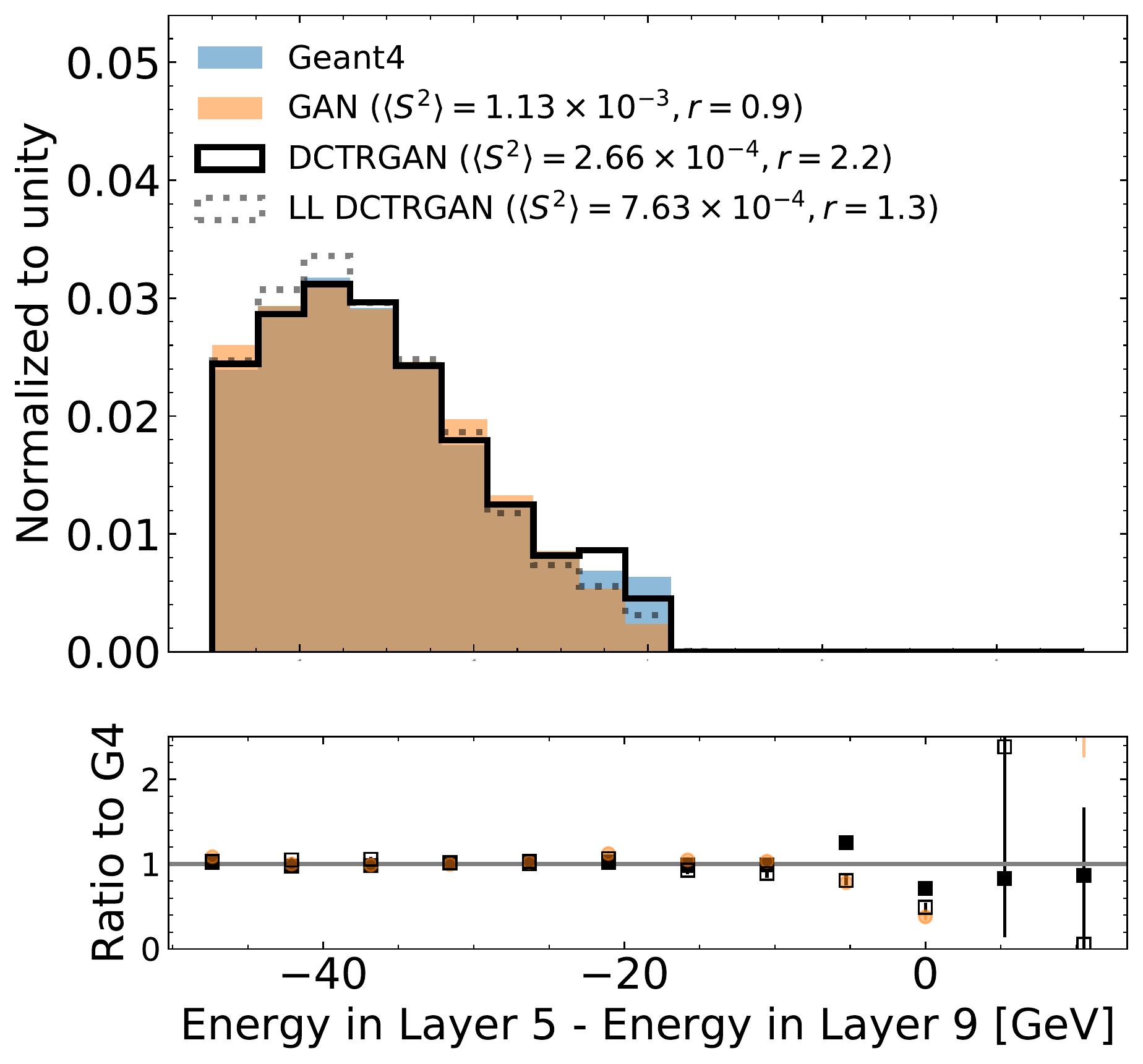}
\includegraphics[height=0.22\textwidth]{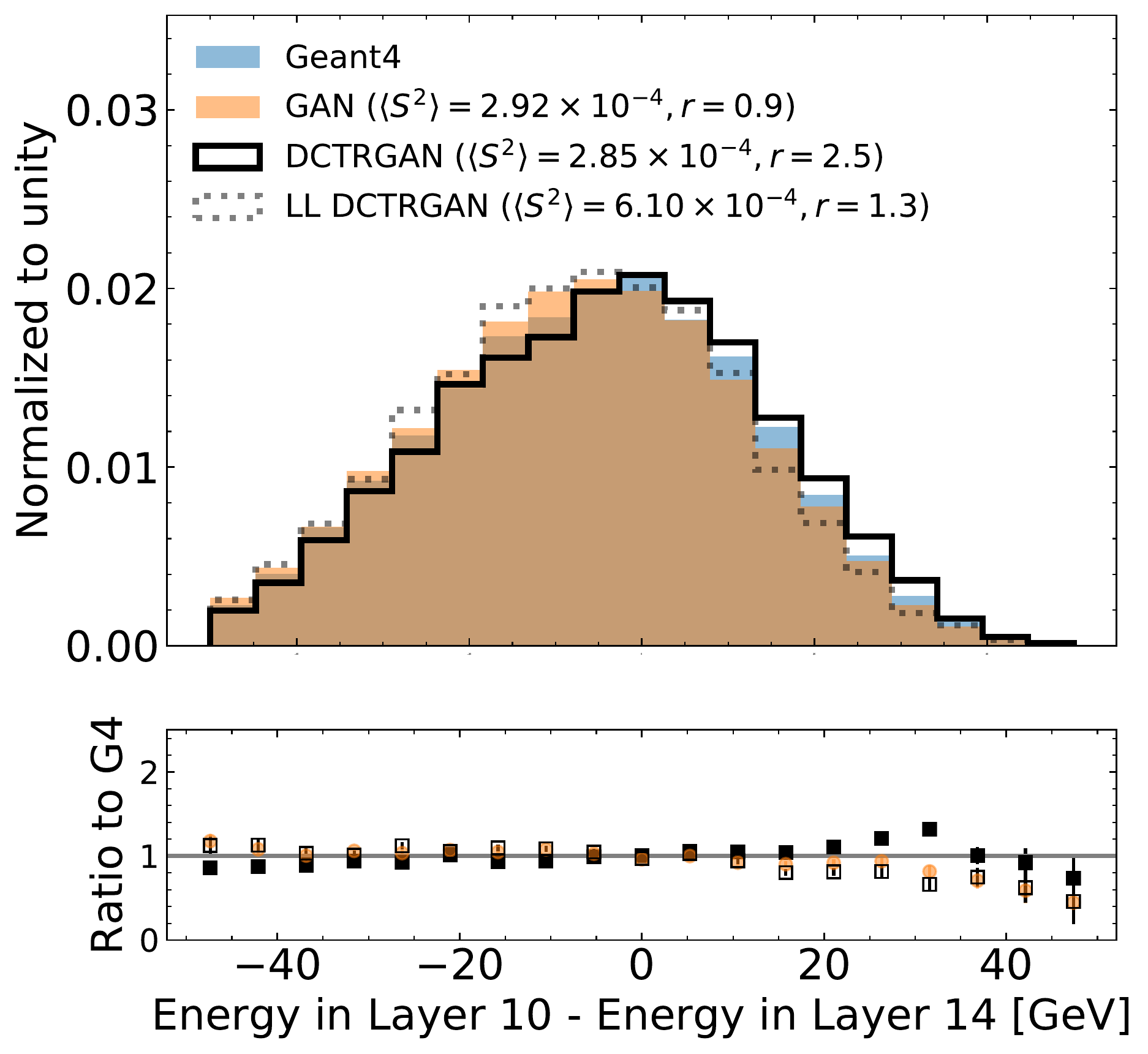}
\includegraphics[height=0.22\textwidth]{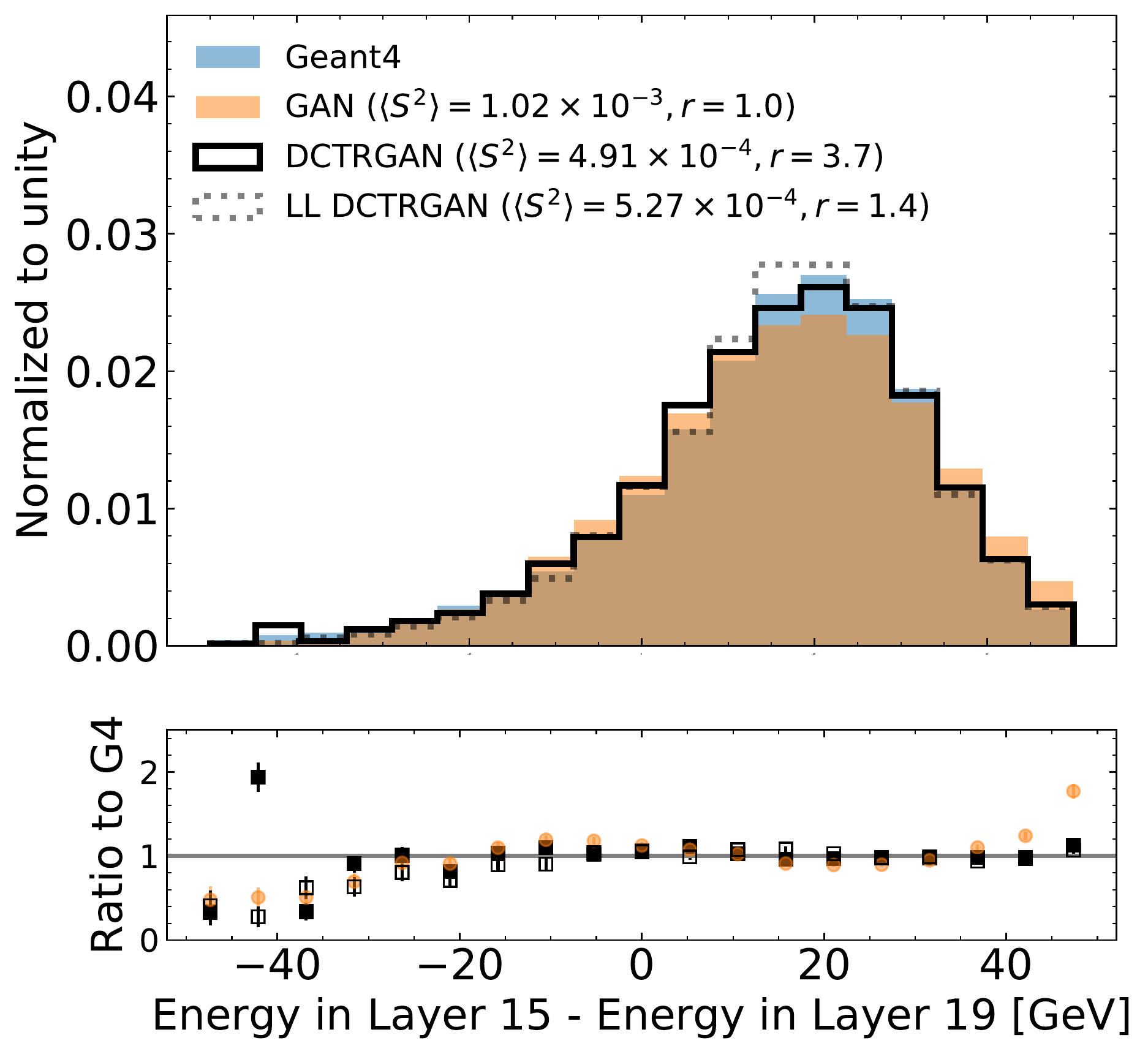}
\includegraphics[height=0.22\textwidth]{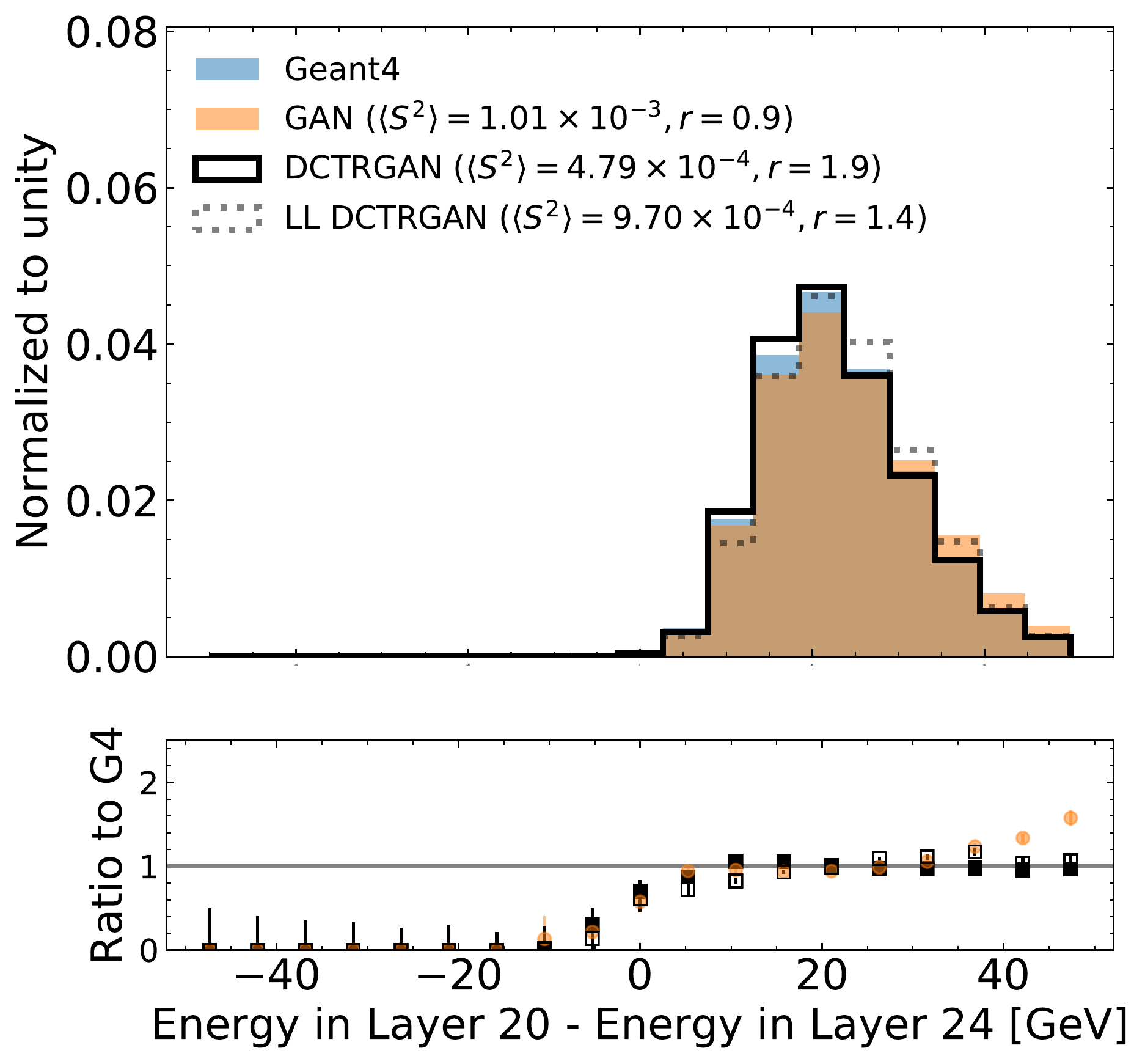}\\
\caption{Selected histograms of total energies differences between layer 5 (left), 10 (second), 15 (third), an 20 (right) with the layer that is one later (top) and four later (bottom). The panels below each histogram show the ratio between the \textsc{Gan} or the \textsc{DctrGan} and the physics-based simulator \textsc{Geant4}.  The legend includes the separation power $\langle S^2\rangle$ between the (weighted) \textsc{Gan} model and the \textsc{Geant}4 model.  Additionally, the ratio $r$ of the uncertainty in the mean of the observable between the \textsc{Gan} and \textsc{Geant}4 is also presented.  Underflow and overflow are not included in the leftmost or rightmost bins.} 
\label{fig:caloexamples4}
\end{figure*}

\begin{figure*}
\centering
\includegraphics[height=0.30\textwidth]{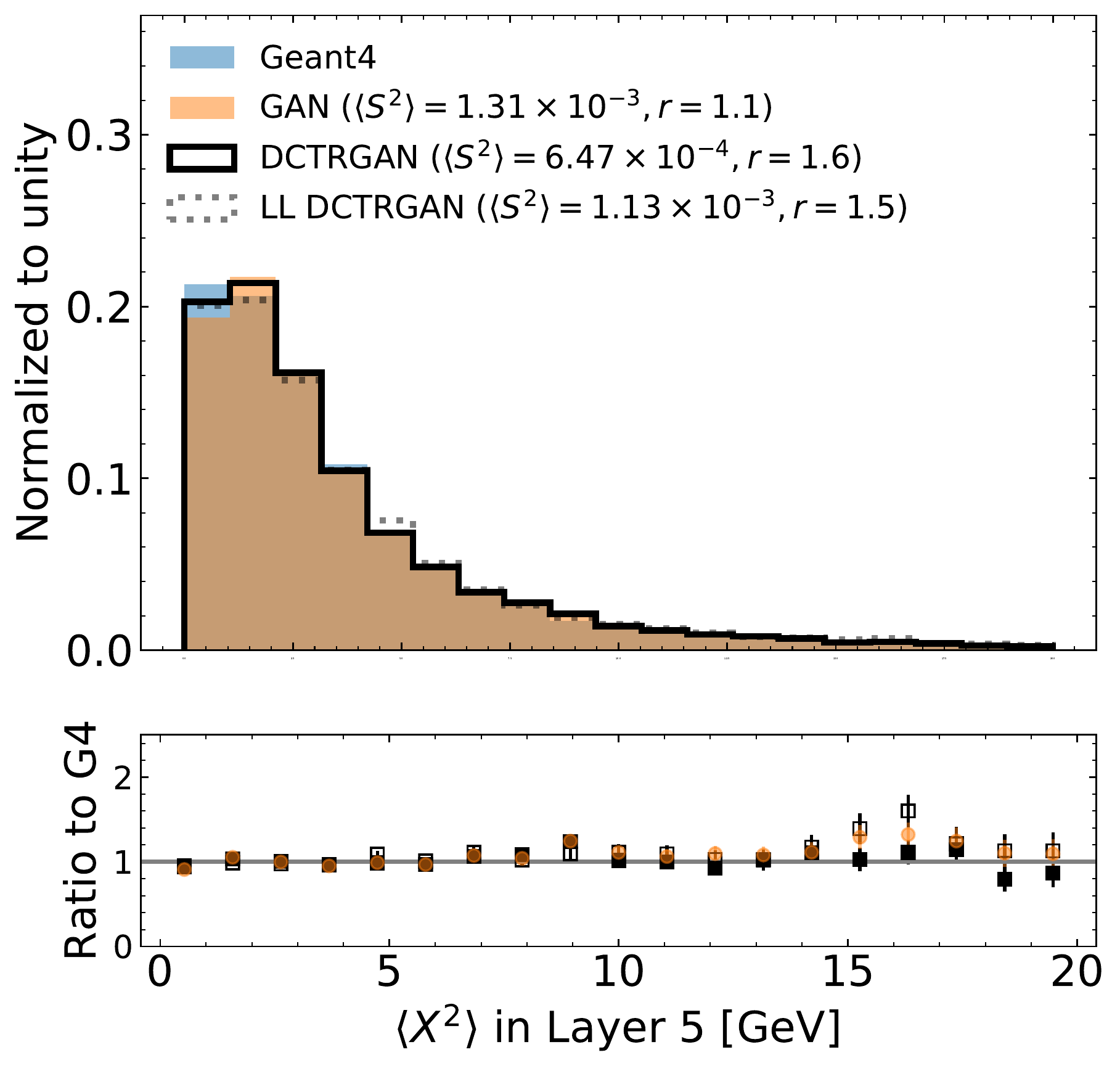}
\includegraphics[height=0.30\textwidth]{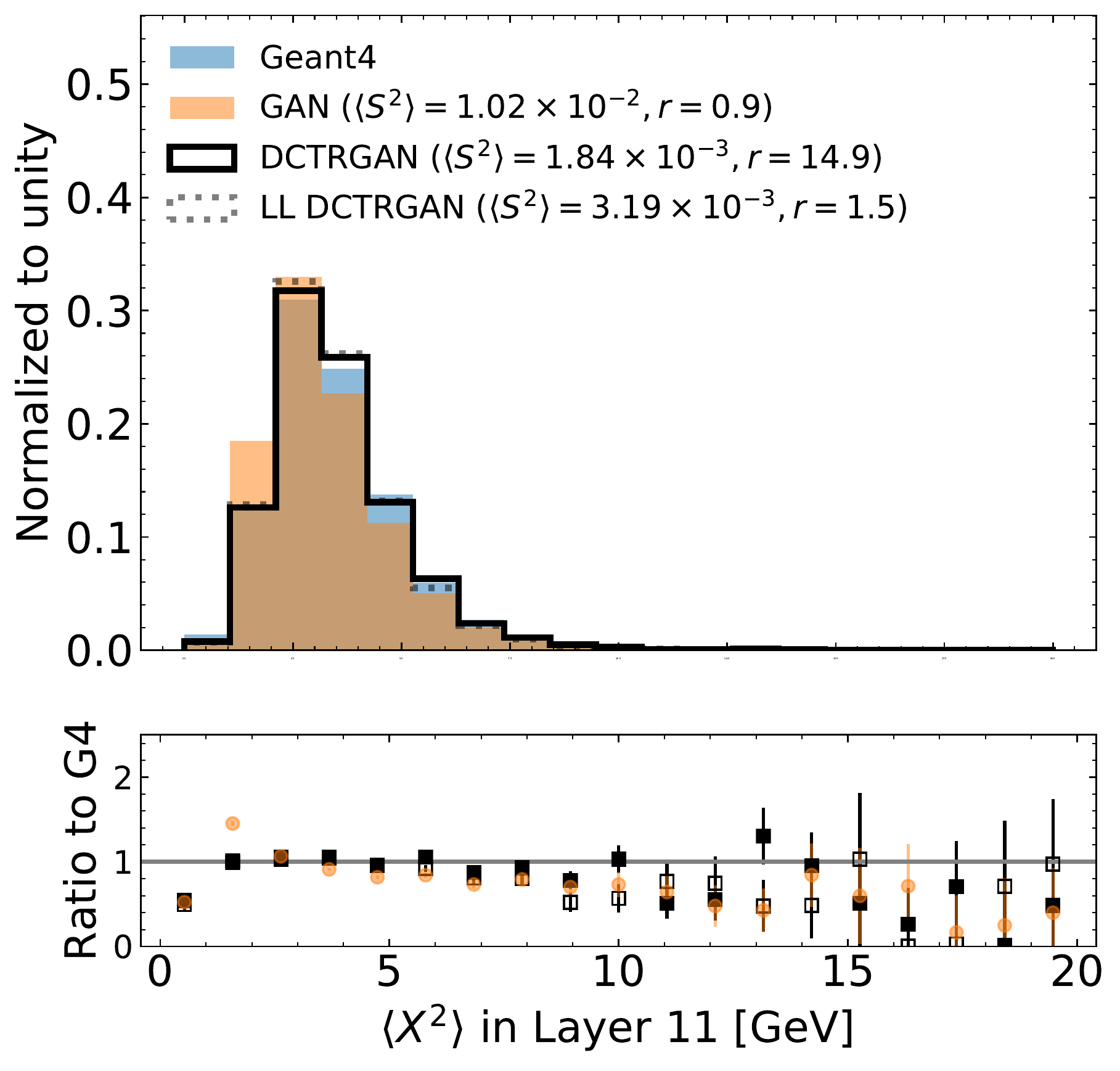}
\includegraphics[height=0.30\textwidth]{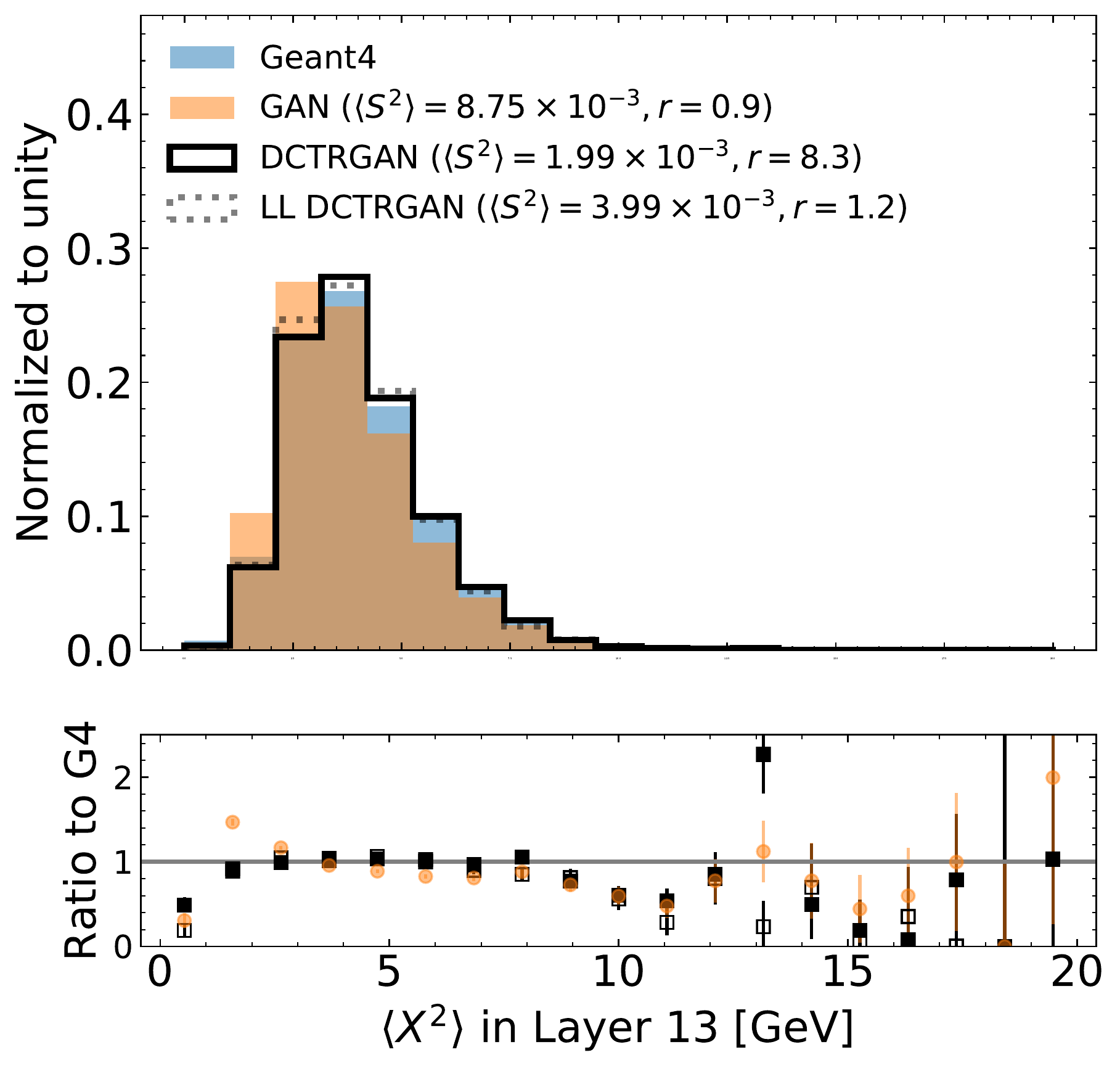} \\
\includegraphics[height=0.30\textwidth]{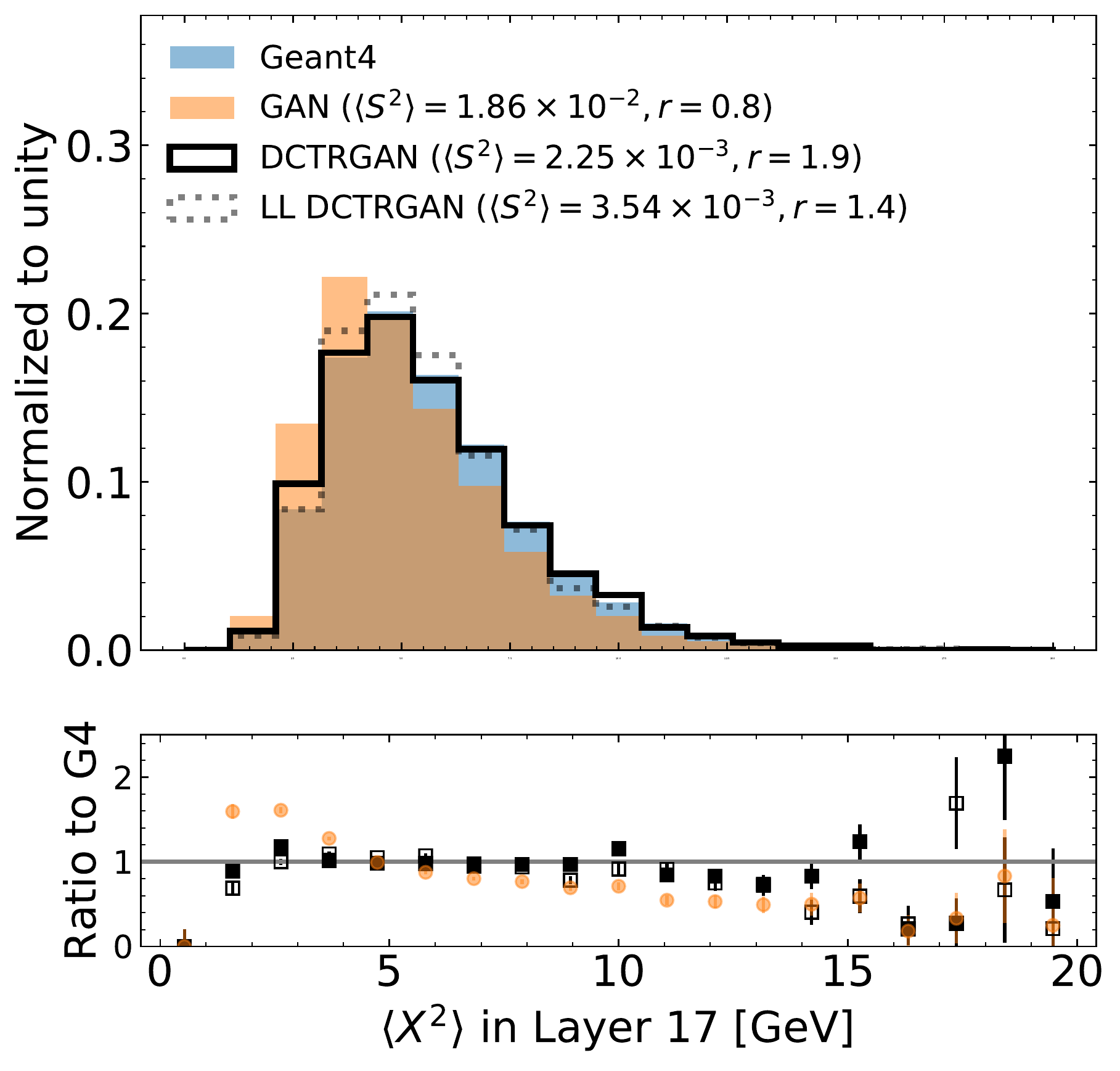}
\includegraphics[height=0.30\textwidth]{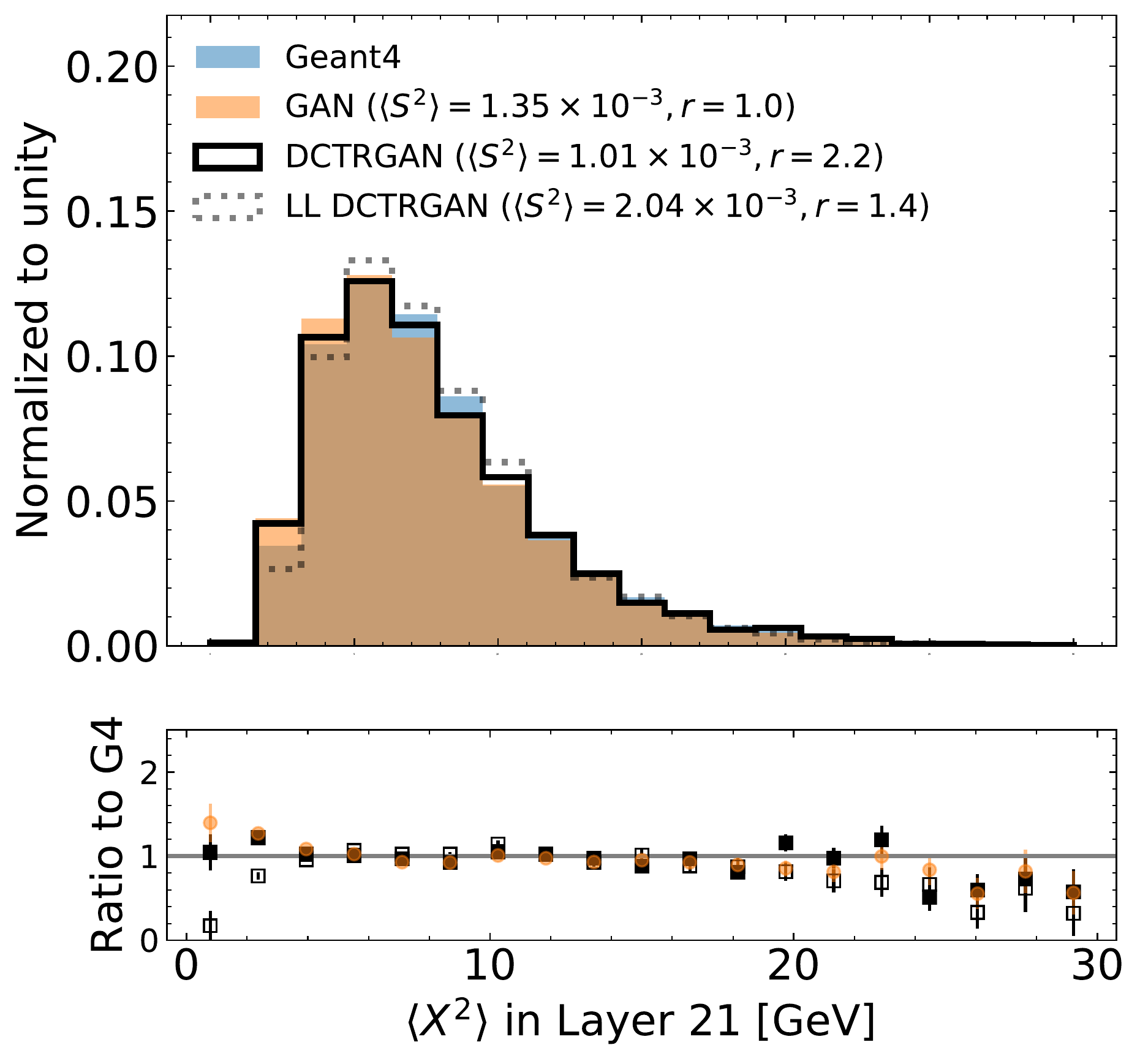}
\includegraphics[height=0.30\textwidth]{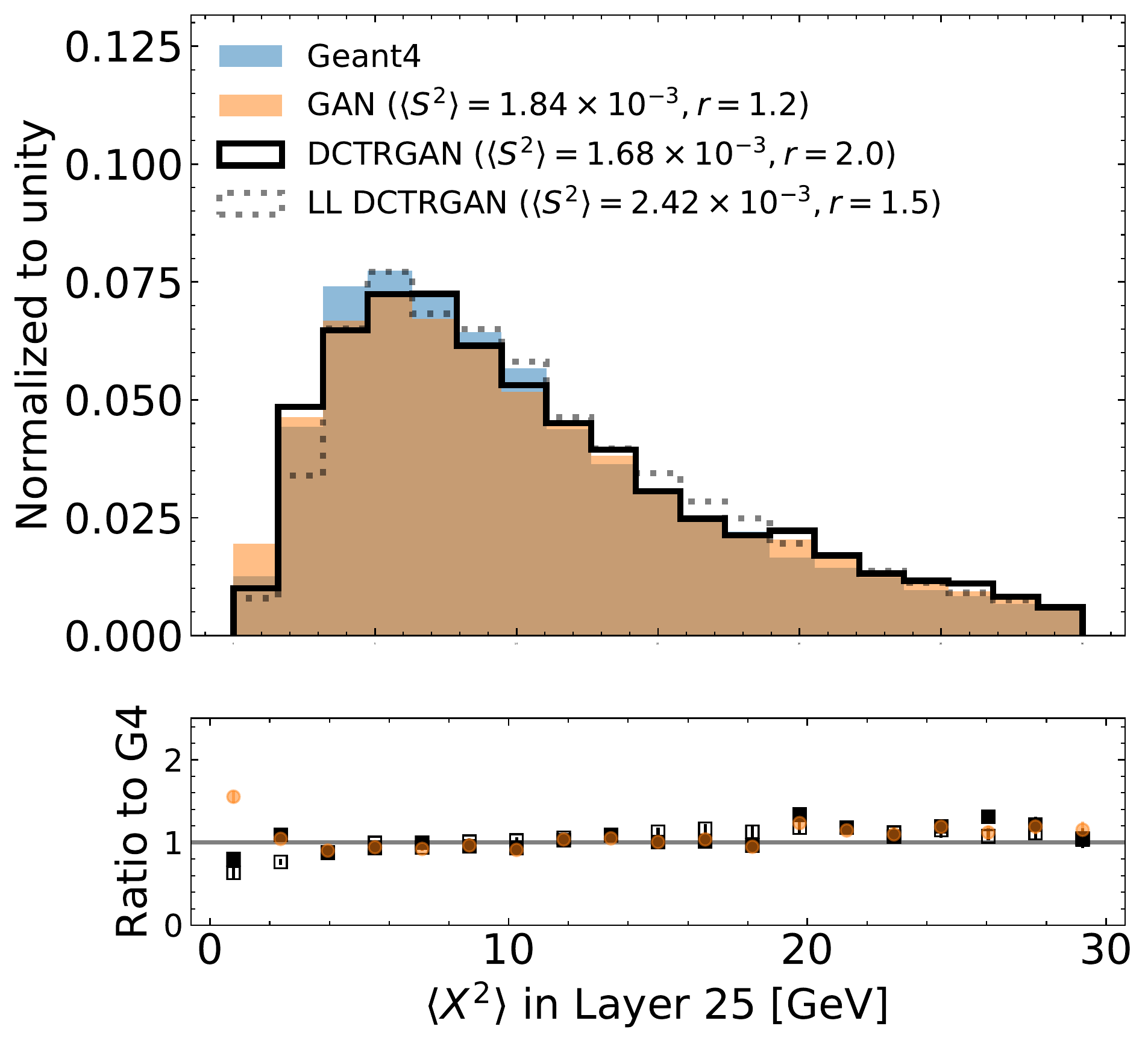}

\caption{Histograms of energy weighted second moment in per layer.  The panels below each histogram show the ratio between the \textsc{Gan} or the \textsc{DctrGan} and the physics-based simulator \textsc{Geant4}.  The legend includes the separation power $\langle S^2\rangle$ between the (weighted) \textsc{Gan} model and the \textsc{Geant}4 model.  Additionally, the ratio $r$ of the uncertainty in the mean of the observable between the \textsc{Gan} and \textsc{Geant}4 is also presented.  Underflow and overflow are not included in the leftmost or rightmost bins.}
\label{fig:caloexamples5}
\end{figure*}

Figures~\ref{fig:caloexamples}-\ref{fig:caloexamples5} show histograms of various one-dimensional observables from the full $30\times30\times30$-dimensional space.  In each plot, two metrics are computed in order to quantify the performance of the generative models.  First, the distance measure
\begin{align}
\langle S^2\rangle=\frac{1}{2}\sum_{i=1}^{n_\text{bins}}\frac{(h_{1,i}-h_{2,i})^2}{h_{1,i}+h_{2,i}}\,,
\end{align}
is used to quantify how similar the \textsc{Gan} histograms are to the \textsc{Geant4} ones.  This measure known in the information theory literature as triangular discrimination~\cite{850703} (and is an $f$-divergence similar to the $\chi^2$ divergence~\cite{Nachman:2016qyc}). In the high-energy physics literature, this is often called the separation power~\cite{Hocker:2007ht}.  It has the property that it is $0$ if the two histograms are the same and 1 when they are non-overlapping in every bin.  As described in Sec.~\ref{sec:weights}, the tradeoff for improved performance is statistical dilution.  To quantify this dilution, the uncertainty on the mean of the observable using the \textsc{Gan} is computed divided by the uncertainty computed using \textsc{Geant4}, denoted $r$ (see Section \ref{sec:weights} for a detailed discussion of these uncertainties).  %
Since the \textsc{Gan} and \textsc{Geant4} datasets have the same number of events, $r\sim 1$ without \textsc{Dctr} and deviates from one for the weighted \textsc{Gan} models.  The effective number of events contained in a weighted sample is observable-dependent, but $r^2$ is approximately by the ratio of the effective number of events in the \textsc{Geant4} dataset to the \textsc{Gan} one.  These values are typically between 1-10 while the \textsc{Gan} is many ten-thousand times faster than \textsc{Geant}4~\cite{Buhmann:2020pmy}.

Three composite observables are presented in Fig.~\ref{fig:caloexamples}.   The total number of activated cells is more peaked around 780 in \textsc{Geant}4 than the \textsc{Gan} and both the low-level and high-level models are able to significantly improve the agreement with \textsc{Geant}4.  The value of $\langle S^2\rangle$ is about 20 times smaller than the unweighted \textsc{Gan} for the high-level \textsc{Dctr} model and about 5 times smaller for the low-level model.  The statistical dilution is modest for the low-level model with $r=1.2$ while it is 3.6 for the high-level model.  The modeling of the total energy is also improved through the reweighting, where both the low-level and high-level models shift the energy towards lower values.  The longitudinal centroid is already relatively well-modeled by the \textsc{Gan}, but is further improved by the high-level \textsc{Dctr} model, reducing the $\langle S^2\rangle$ by more than a factor of two.

Histograms of the energy in representative layers are shown in Fig.~\ref{fig:caloexamples2}.  Generally, the \textsc{Geant}4 showers penetrate deeper into the calorimeter than the \textsc{Gan} showers, so the energy in the early layers is too high for the \textsc{Gan} and the energy in the later layers is too low.  The \textsc{Dctr} models are able to correct these trends, with a systematically superior fidelity as measured by $\langle S^2\rangle$ for the high-level model.  

The modeling of correlations between layers is probed in Fig.~\ref{fig:caloexamples4} by examining histograms of the difference between energies in different layers.  Layers that are closer together should be more correlated.  This manifests as histograms with a smaller spread for layers that are physically closer together.  For layers that intercept the shower before its maximum, the difference in energy between a layer and the next layer is negative.    The shower maximum is typically just beyond the tenth layer.  The fidelity improvements from the weighting in the difference histograms is comparable to the per-layer energy histograms from Fig.~\ref{fig:caloexamples2}. Interestingly, in cases when the $\langle S^2\rangle$ for the \textsc{Gan} is already good without weighting (e.g. the energy in layer 15 - layer 16), the modeling still improves with the \textsc{Dctr} approach and there can still be significant statistical dilution.  

\begin{figure*}
\centering
\includegraphics[height=0.32\textwidth]{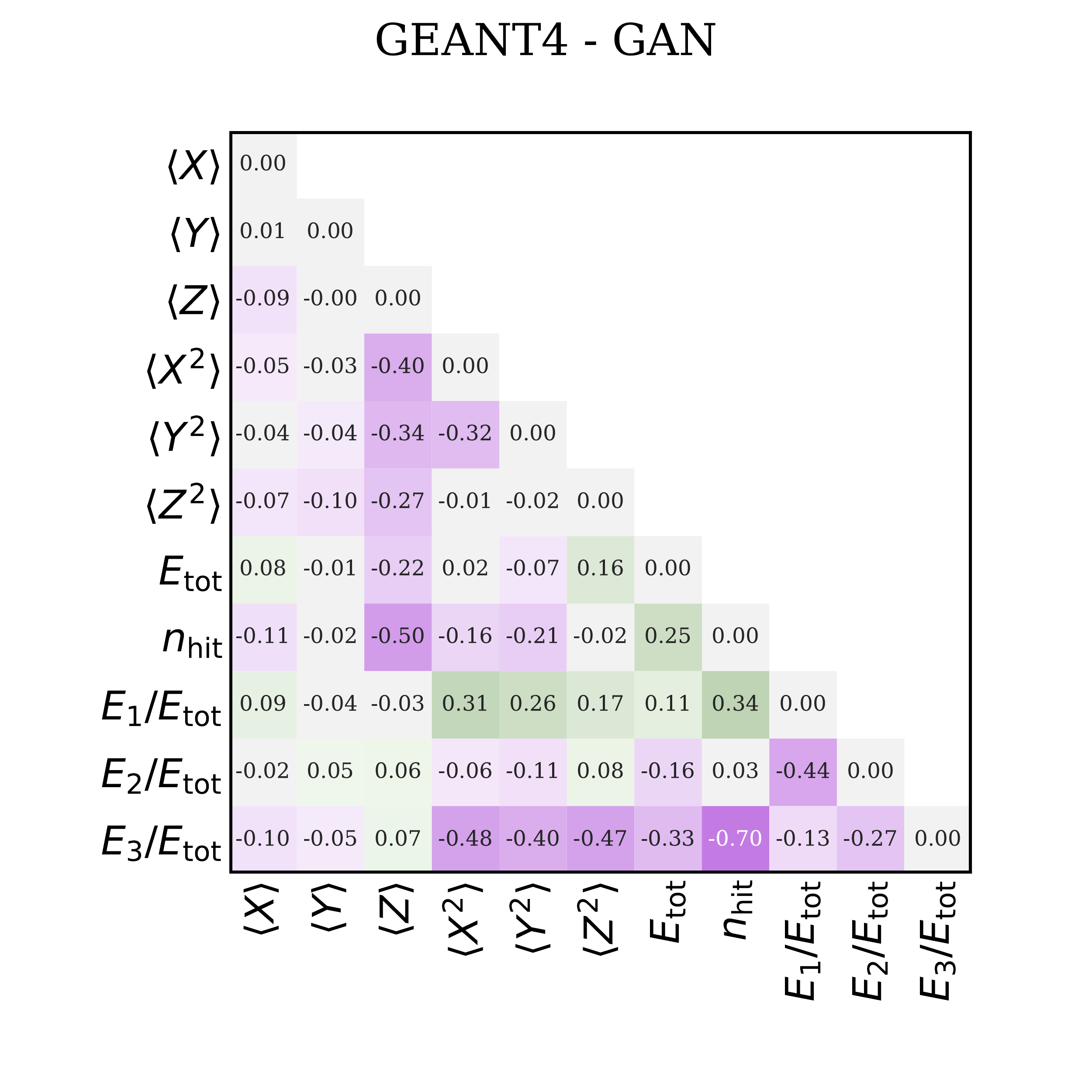}\includegraphics[height=0.32\textwidth]{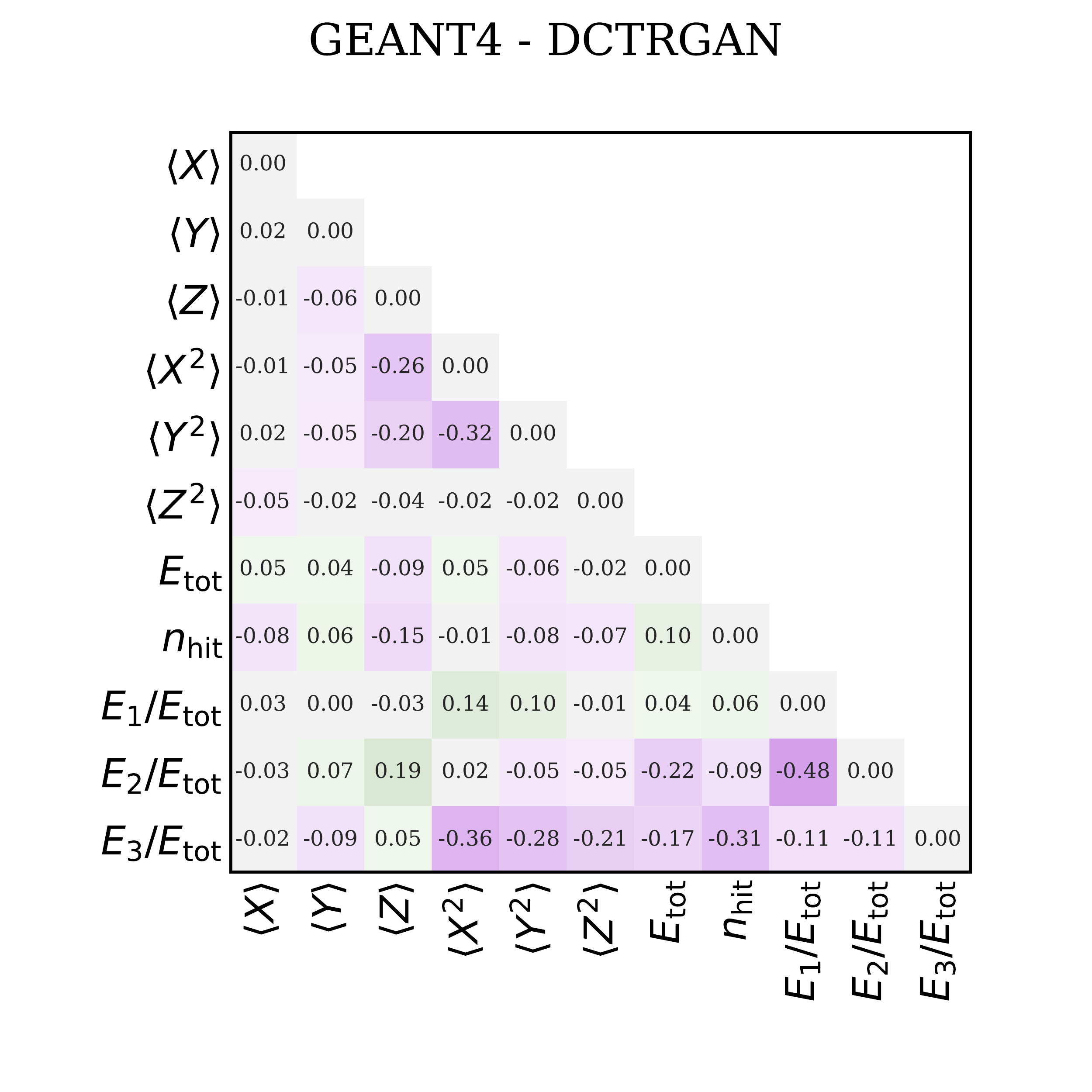}\includegraphics[height=0.32\textwidth]{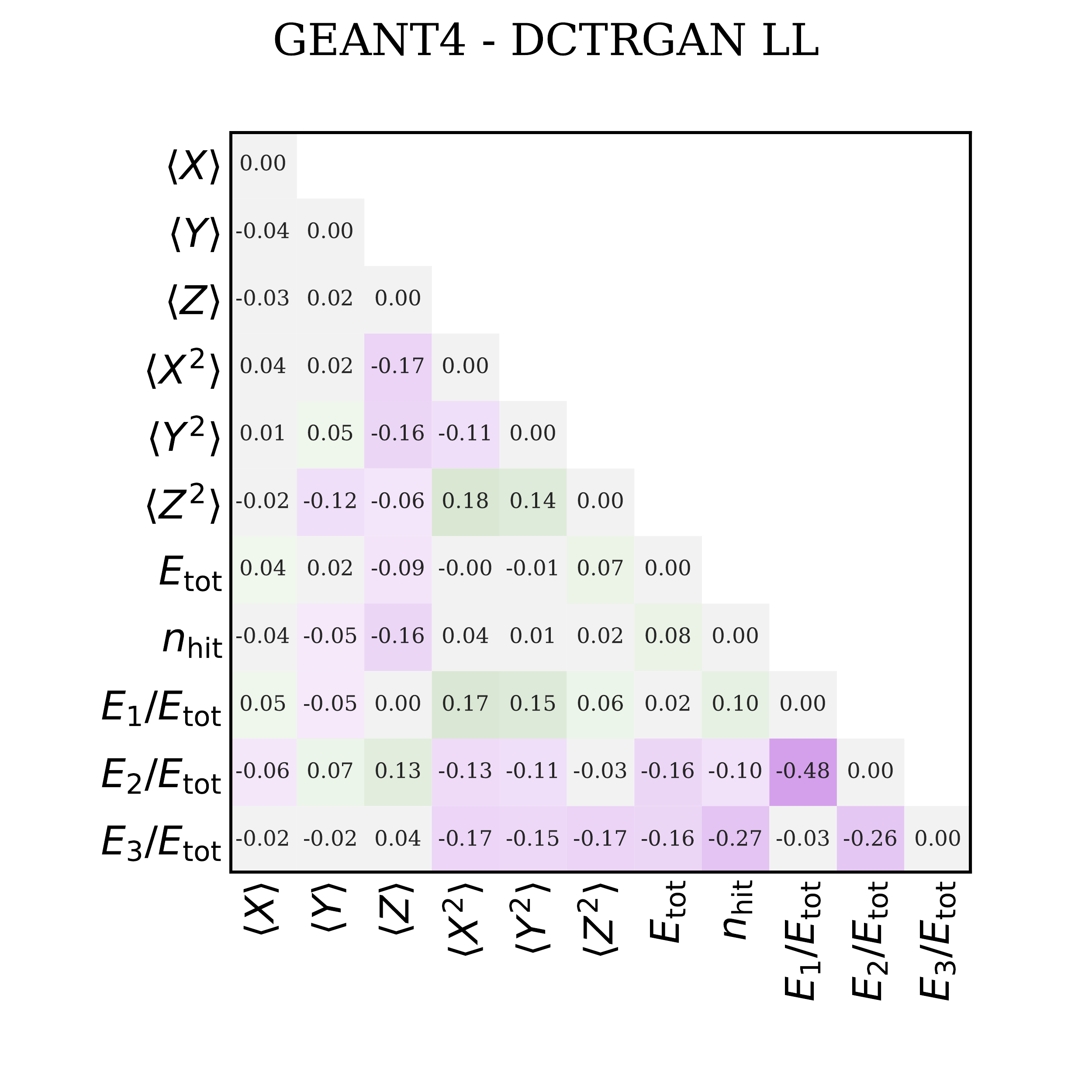}\\

\caption{Differnces in pairwise linear correlations between the stated observables between \textsc{Geant}4 and the nominal \textsc{Gan} (left), for the high-level \textsc{DctrGan} (middle) and the low-level \textsc{DctrGan} (right).  Darker colors indicate stronger correlation.}
\label{fig:correlations}
\end{figure*}

Another way to visualize correlations in the data is to compute the linear correlation coefficient between observables.  This is reported for a representative set of observables in Fig.~\ref{fig:correlations}.  Generally, the difference in correlations between the \textsc{Gan} and \textsc{Geant}4 are improved after applying the \textsc{Dctr} reweighting, with many of the residual differences reducing by factors of about 2-10.

Figures~\ref{fig:caloexamples}-\ref{fig:caloexamples4} are features that were directly used by the high-level model.  Figure~\ref{fig:caloexamples5} presents histograms for a collection of observables that were not accessible to the high-level model during training.   In particular, the energy-weighted second moment along the $x$-direction are computed for each layer.  The results for the $y$-direction are nearly the same.  Despite not being present during training, the high-level network is still able to improve the the performance over the unweighted \textsc{Gan} in every case with only a modest reduction in statistical power.   Weights in the \textsc{Dctr} models are per-example so one can compute any observable even if they are not explicitly part of the reweighting model evaluation.

\begin{figure*}
\centering
\includegraphics[height=0.22\textwidth]{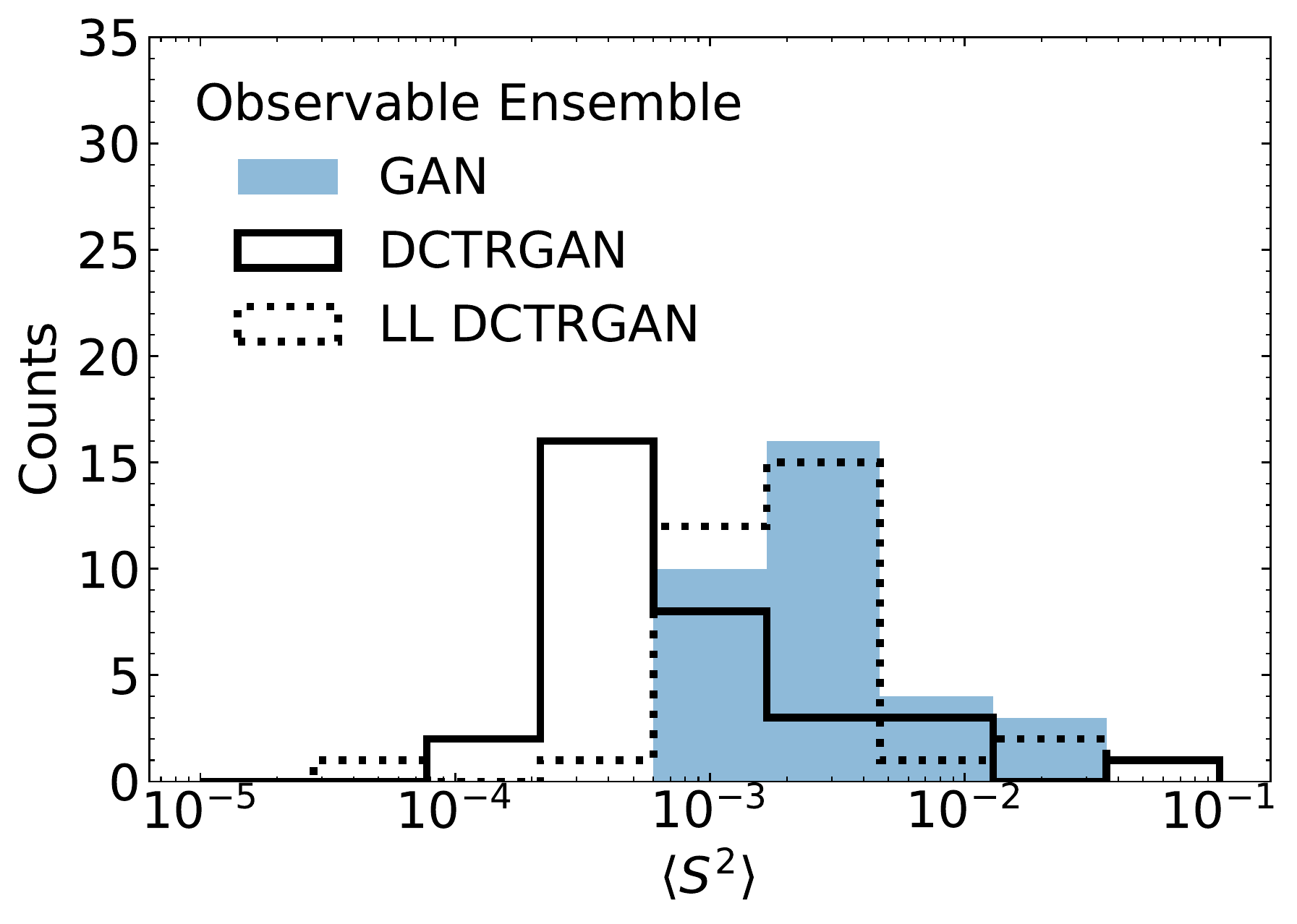}\includegraphics[height=0.22\textwidth]{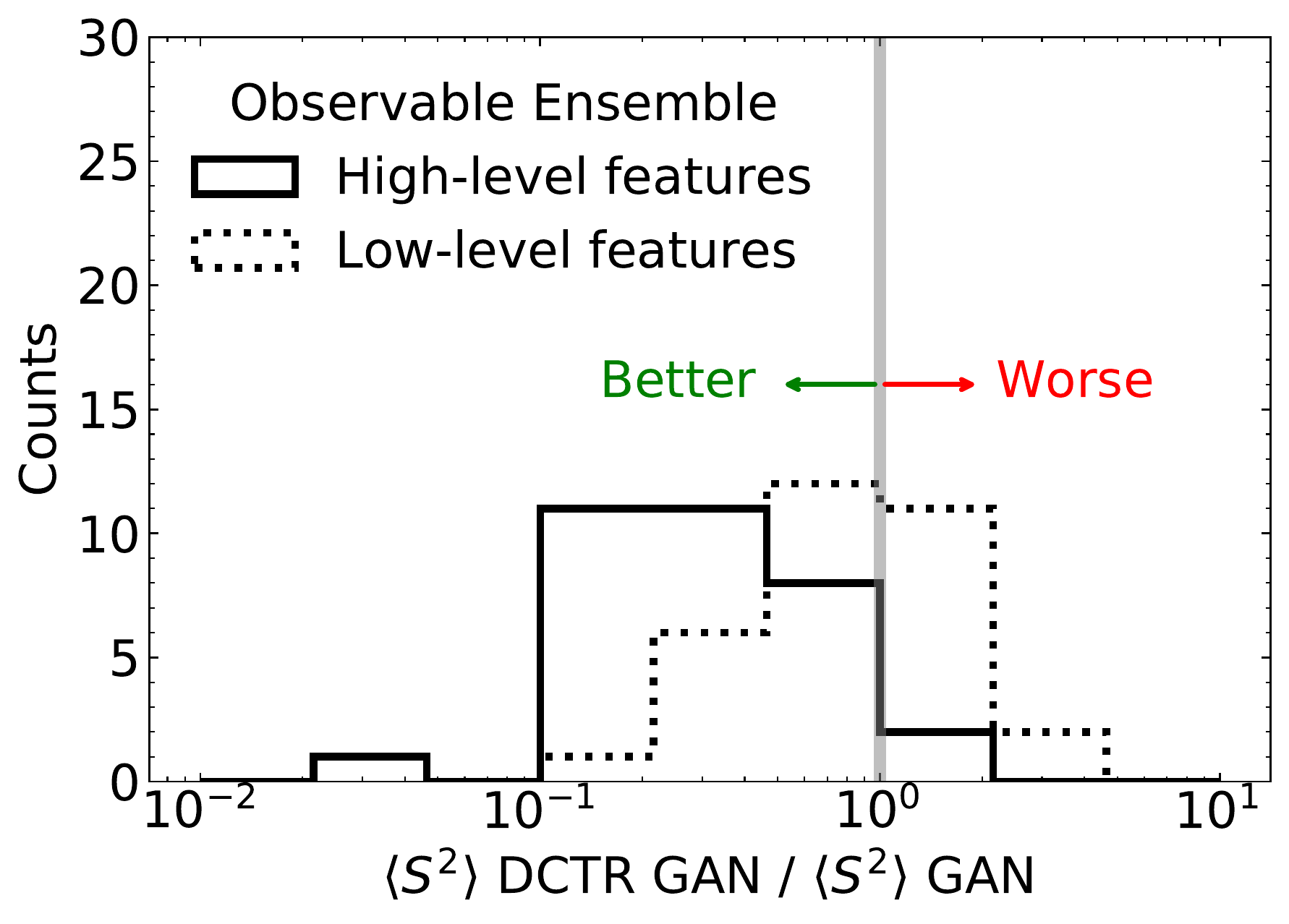}\includegraphics[height=0.22\textwidth]{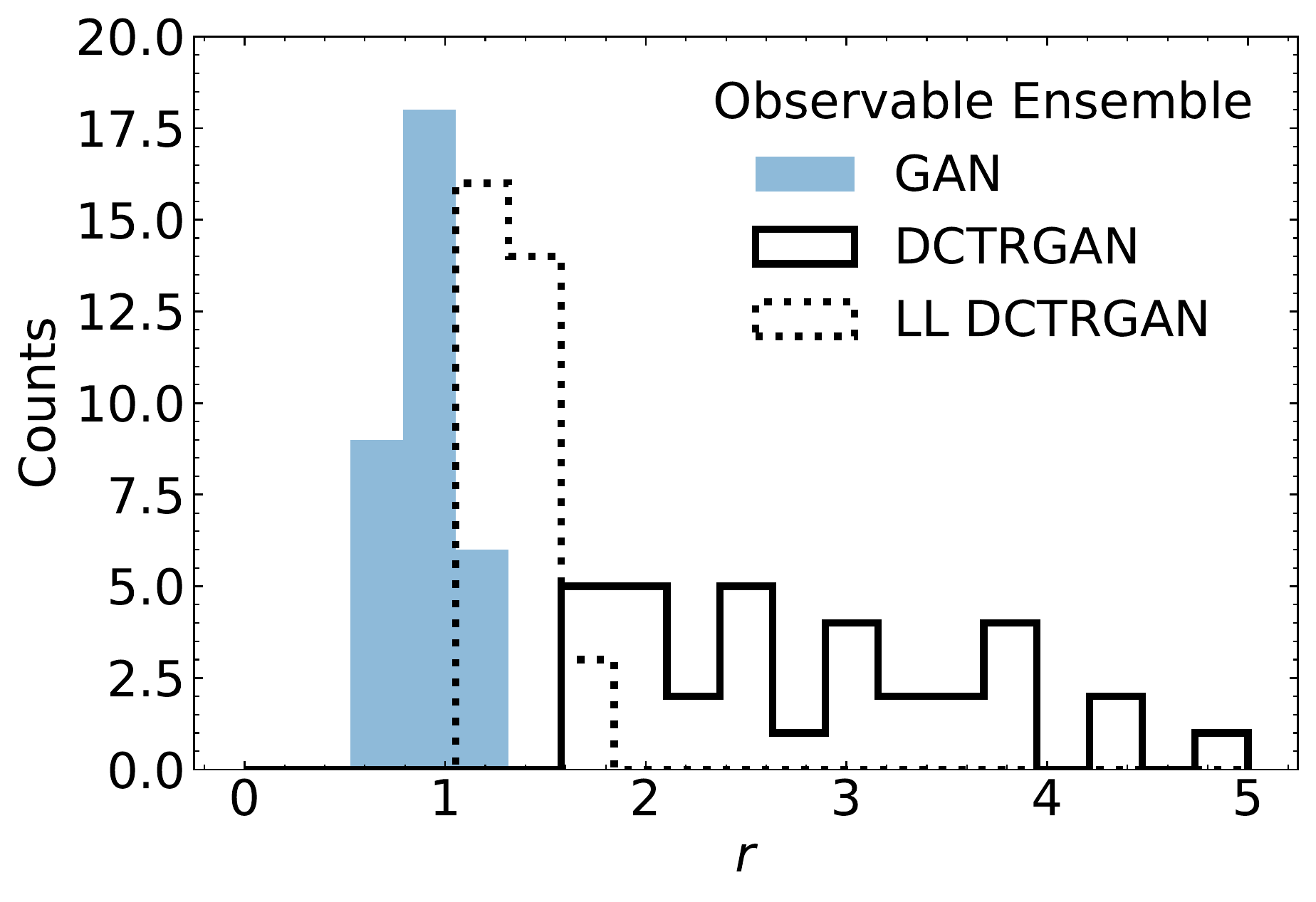}\\
\includegraphics[height=0.22\textwidth]{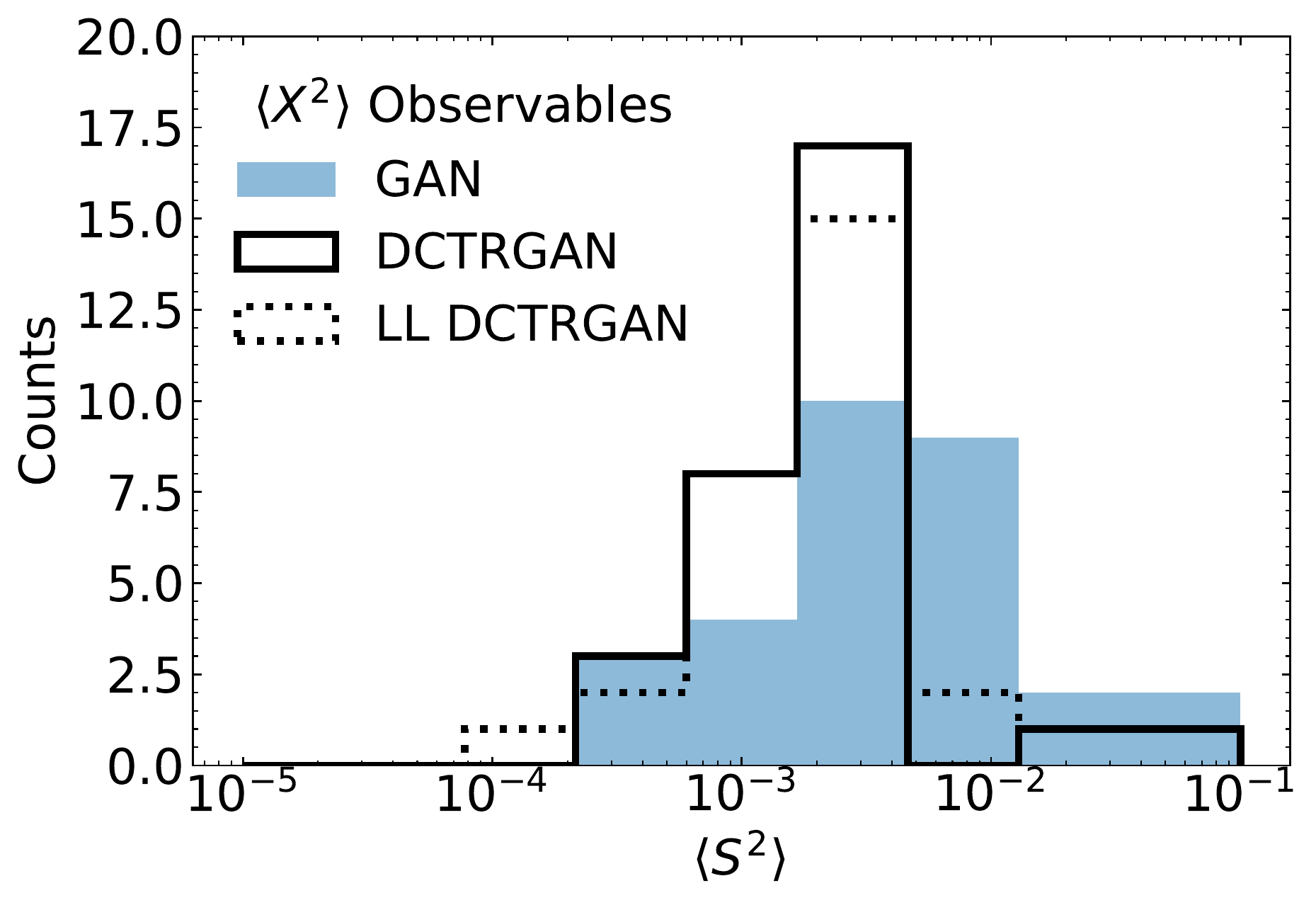}\includegraphics[height=0.22\textwidth]{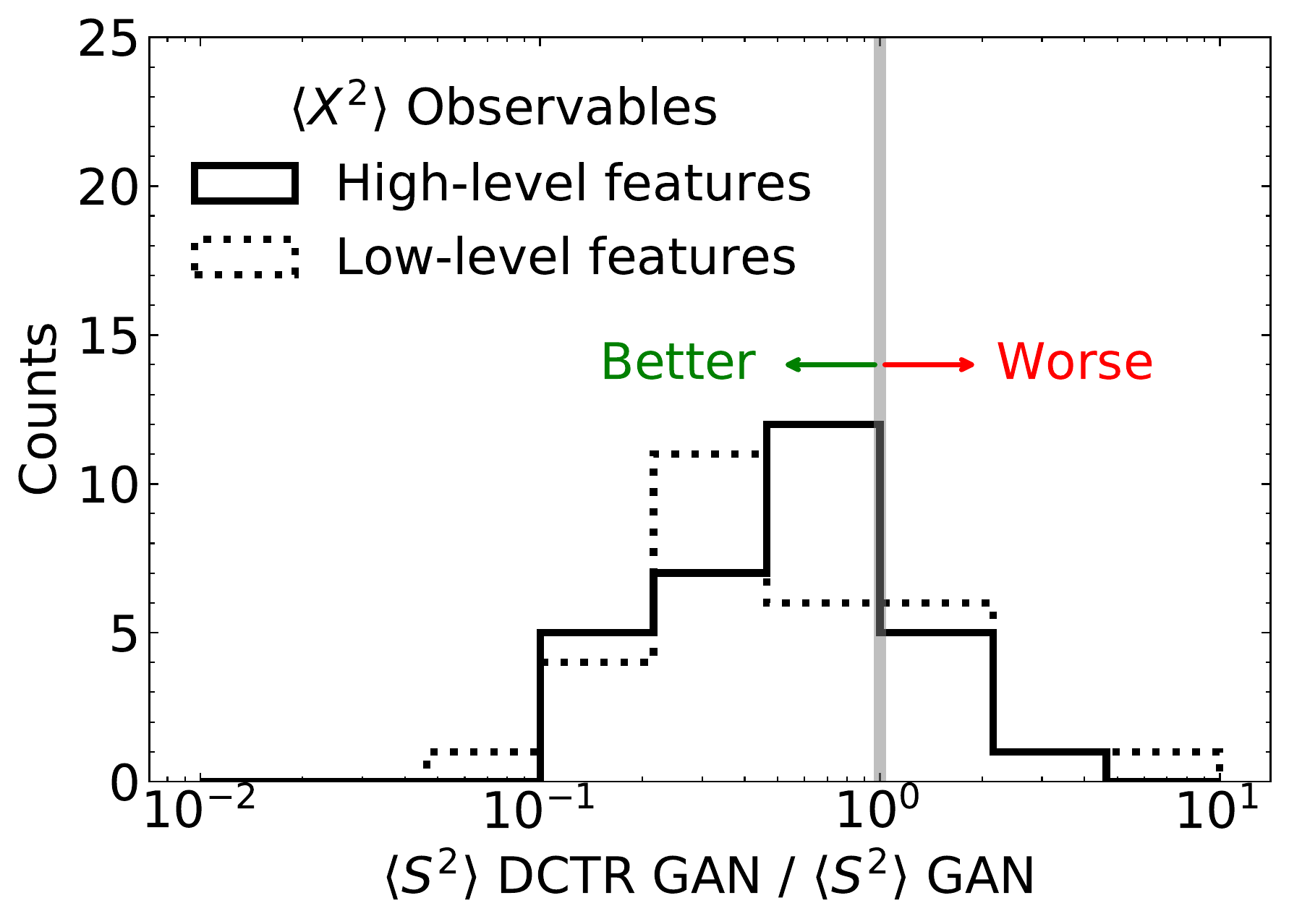}\includegraphics[height=0.22\textwidth]{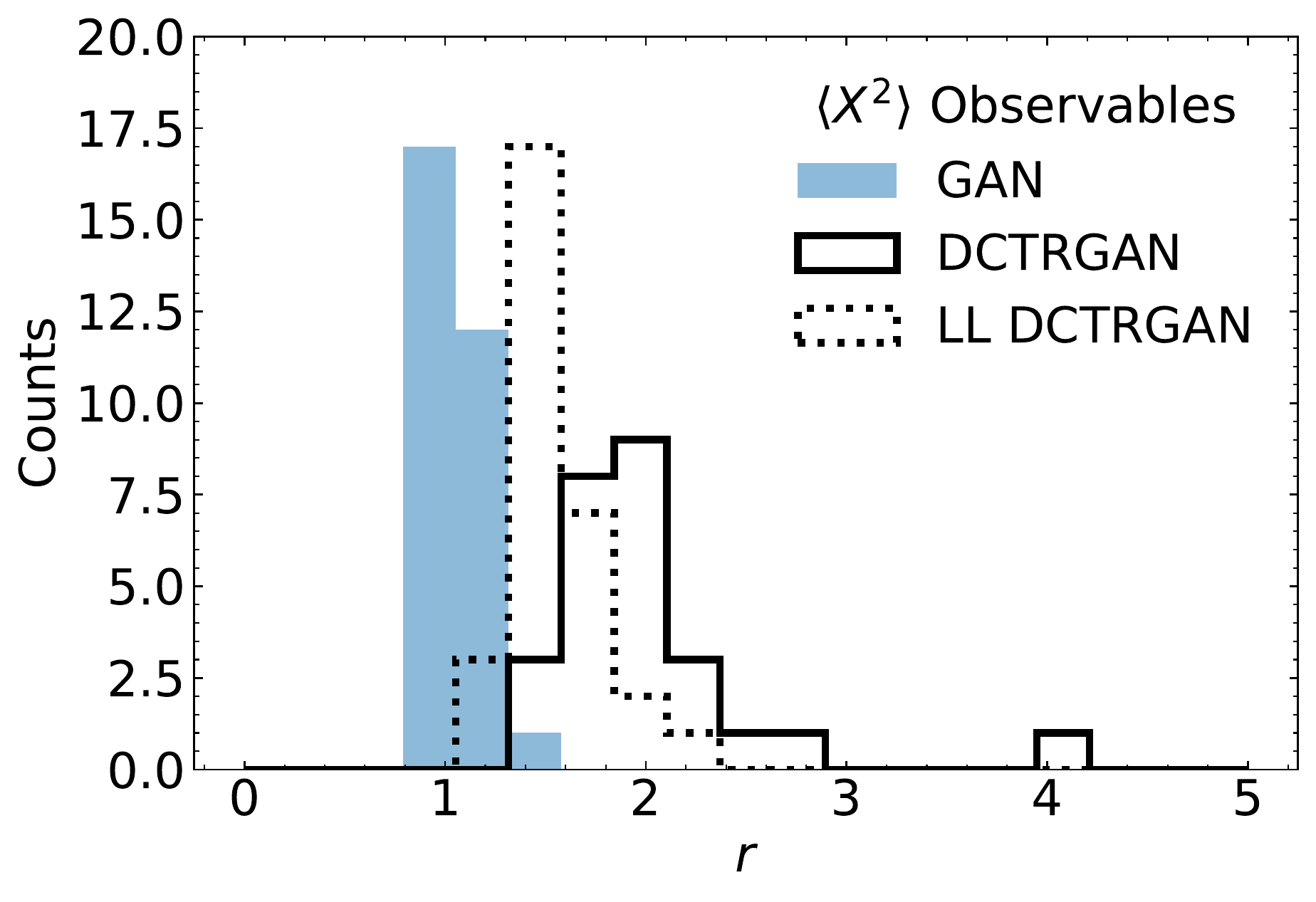}\\
\includegraphics[height=0.22\textwidth]{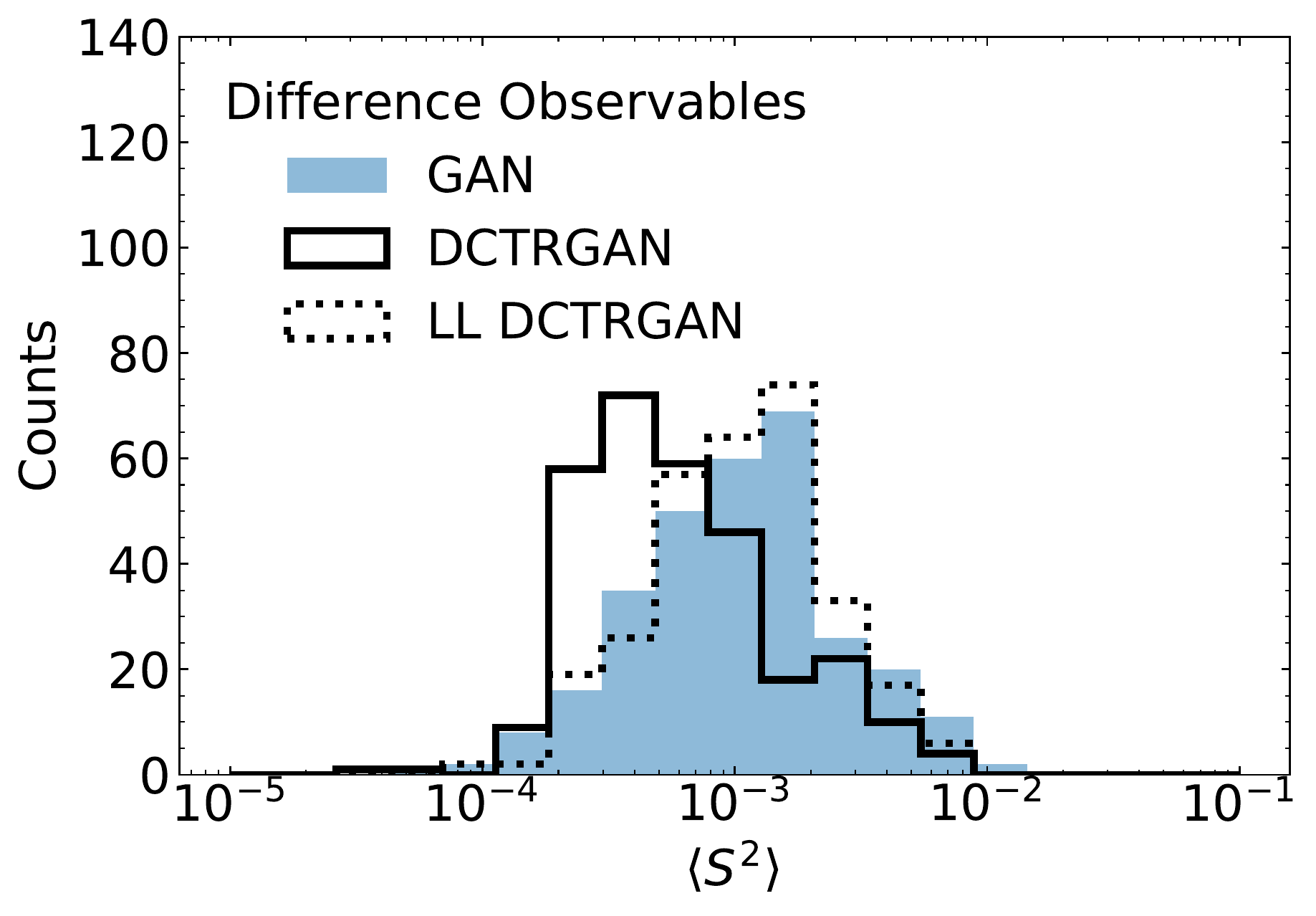}\includegraphics[height=0.22\textwidth]{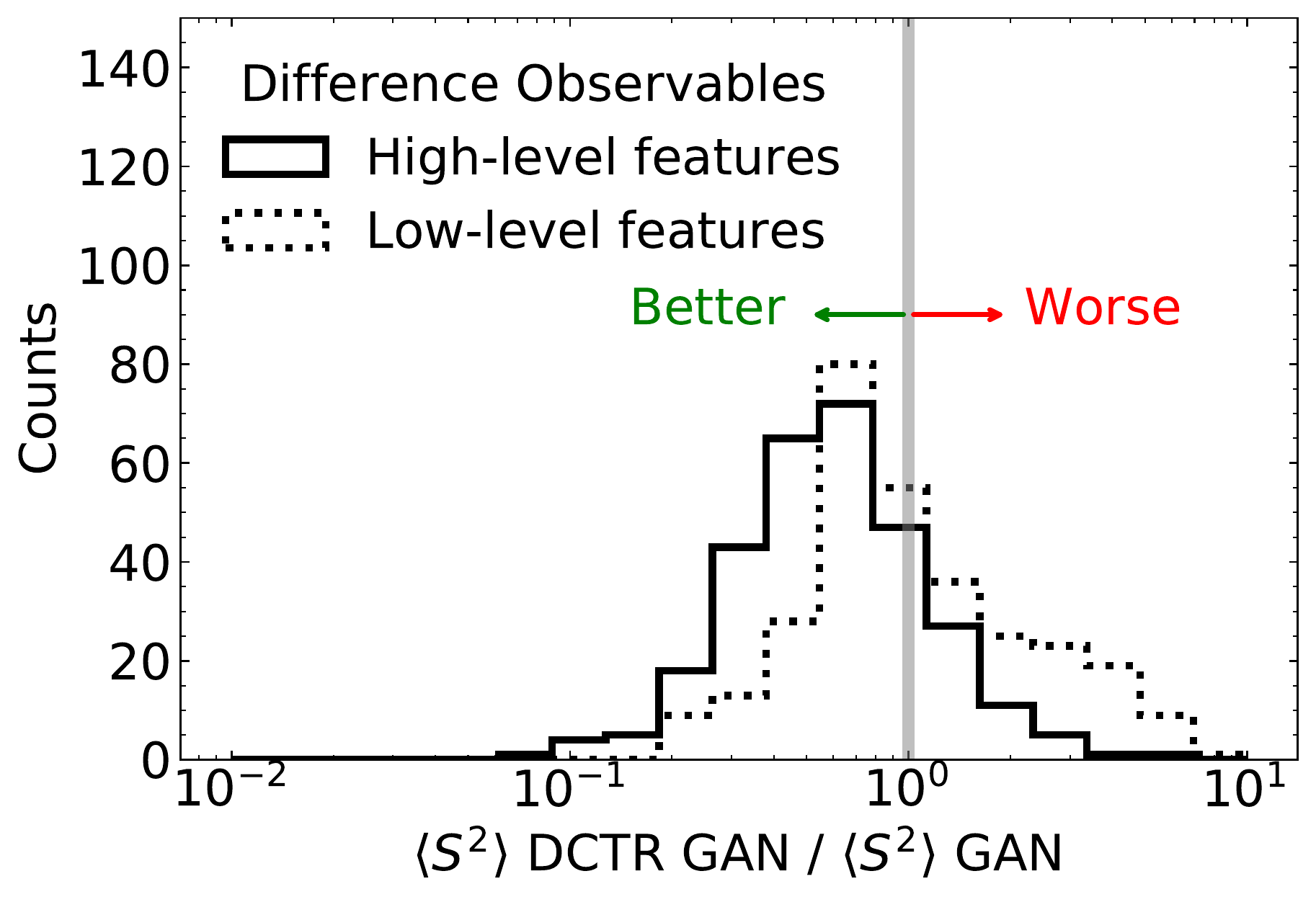}\includegraphics[height=0.22\textwidth]{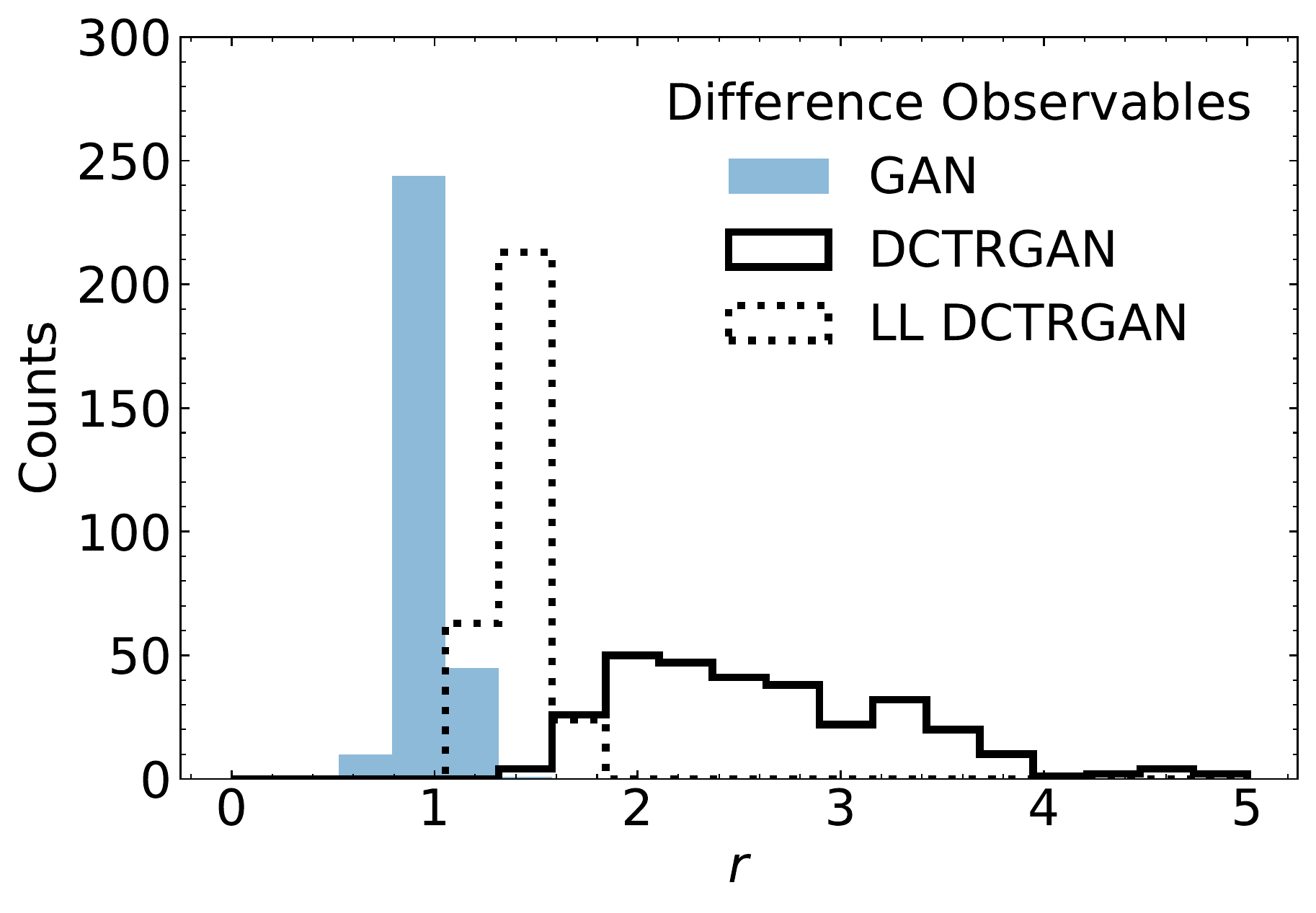}\\
\caption{Histograms that summarize the performance of the reweighting models.  The first column presents histograms of $\langle S^2\rangle$ for all of the observables shown in the previous figures and the second column shows the ratio of $\langle S^2\rangle$ for the \textsc{Gan} compared with the \textsc{DctrGan} models.  The third column quantifies the statistical dilution using $r$.  The top row includes all of the 33 input features to the high-level network, the second row includes all 30 $\langle X^2\rangle$ observables, and the third row includes all differences between layers. }
\label{fig:overview}
\end{figure*}

A summary of the overall performance appears in Fig.~\ref{fig:overview}.  The most probable ratio of $\langle S^2\rangle$ computed with \textsc{DctrGan} to the one computed with the unweighted \textsc{Gan} is between 0.4 and 0.5 and most most of the observables show improvement after weighting.  As mentioned earlier, the \textsc{Gan} and \textsc{Geant4} datasets have the same number of events so $r\sim 1$ without \textsc{Dctr}.  For the low-level model, $r$ has a narrow distribution peaked at about 1.5.  In contrast, the high-level model is peaked past two with a tail out to around 5.  This difference in $r$ and the overall performance between the low- and high-level models is in part from the extensive regularization of the low-level model during training.  The high-level model is also highly-regularized by the dimensionality reduction, but otherwise has a sufficiently complex classifier that is not arrested during training.

\clearpage

\section{Conclusions and Outlook}
\label{sec:conclusions}

This paper has introduced a post-processing method for generative models to improve their accuracy.  
While the focus lies on deep generative models, it could similarly be used to enhance
the precision of other fast simulation tools (such as \textsc{Delphes}~\cite{deFavereau:2013fsa}).
The approach is based on reweighting and the result is a combined generator and weighting function that produces weighted examples.  These weighted examples can be used to perform inference using the same tools as for unweighted events.  The potential of deep generative models continues to expand as their fidelity improves and tools like \textsc{DctrGan} may be able help achieve the target precision for a variety of applications.

\section*{Code and Data}

The code for this paper can be found at \url{https://github.com/bnachman/DCTRGAN}.  Examples and instructions to reproduce the calorimeter GAN dataset
are available at \url{https://github.com/FLC-QU-hep/getting_high}.

\section*{Acknowledgments}

We thank A. Andreassen, P. Komiske, E. Metodiev, and J. Thaler for many helpful discussions about reweighting with NNs and E. Buhmann, F. Gaede,  and K. Kr\"{u}ger for stimulating discussions on improving the fidelity of calorimter simulation.  We also thank J. Thaler for feedback on the manuscript.  BPN and DS are supported by the U.S.~Department of Energy, Office of Science under contract numbers DE-AC02-05CH11231 and DE-SC0010008, respectively.  BPN would also like to thank NVIDIA for providing Volta GPUs for neural network training. GK and SD acknowledge support by the DFG under Germany’s Excellence Strategy – EXC 2121 \textsl{Quantum Universe – 390833306}. EE is funded through the Helmholtz Innovation Pool project AMALEA that provided a stimulating scientific environment for parts of the research done here. DS is grateful to LBNL, BCTP and BCCP for their generous support and hospitality during his sabbatical year.

\bibliography{HEPML,myrefs}

\end{document}